\documentclass[]{tMOP2e}
\usepackage[ansinew]{inputenc}
\usepackage[T1]{fontenc}
\usepackage[english]{babel}
\usepackage{amsfonts}
\usepackage{amsmath}
\usepackage{color}
\newcommand{\h}{\mathcal{H}}
\newcommand{\beq}{\begin{equation}}
\newcommand{\eeq}{\end{equation}}
\newcommand{\ran}{\rangle}
\newcommand{\lan}{\langle}
\newcommand{\Tr}{\textrm{Tr}}
\newcommand{\N}{\mathbb{N}}
\newcommand{\C}{\mathbb{C}}
\newcommand{\M}{\mathcal{M}}
\newcommand{\cald}{\mathcal{D}}
\newcommand{\R}{\mathbb{R}}
\newcommand{\To}{\longrightarrow}
\newcommand{\ii}{\textrm{i}}
\newcommand{\pr}{^{\prime}}
\newcommand{\und}{\underline}
\newcommand{\0}{\mathbb{O}}

\newcommand{\norm}[1]{\left\Vert#1\right\Vert}

\title{{\itshape Parametrizations of density matrices}}
\author{E. Br\"uning$^a$$^\ast$\thanks{$^\ast$Corresponding author. Email: bruninge@ukzn.ac.za \vspace{6pt}},
H. M\"akel\"a$^{c,d}$, and A. Messina$^d$, F. Petruccione$^b$ \\ \vspace{6pt}
$^a${\em{School of Mathematical Sciences, University of KwaZulu-Natal, Durban, South Africa}};
 $^{b}${\em{School of Physics, University of KwaZulu-Natal, Durban, South Africa}};
$^c${\em{Department of Physics, Ume\aa \,\, University, Ume\aa, Sweden}};
 $^{d}${\em{Department of Physics, University of Palermo, Palermo, Italy}}\\
 \vspace{6pt}
 \received{August 2011}
 }

\begin{document}

\maketitle
\begin{abstract}
This article gives a brief overview of some recent progress in the characterization and parametrization of density matrices of finite dimensional systems. We discuss in some detail the Bloch-vector and Jarlskog parametrizations and mention briefly the coset parametrization. As applications of the Bloch parametrization we discuss the trace invariants for the case of time dependent Hamiltonians and in some detail the dynamics of three-level systems. Furthermore, the Bloch vector of two-qubit systems as well as the use of the polarization operator basis is indicated. As the main application of the Jarlskog parametrization we construct density matrices for composite systems. In addition, some recent related articles are mentioned without further discussion.
\end{abstract}

\section{Introduction}

\subsection{Motivation}
The state of a quantum system can be mathematically represented as a density matrix,
a positive semidefinite hermitian operator of trace one.
The density matrices of finite dimensional systems, which are the topic of this review, can be expressed as complex $n\times n$ matrices, constrained by the hermiticity, positivity, and trace conditions.
Considering the fundamental role of density matrices for the description of physical systems, it is not surprising that many investigations have been devoted to the parametrizations of density matrices over the last years. This work is motivated by many reasons.
For example, a suitably chosen parametrization may considerably simplify solving a specific physical
problem, it may help to identify new properties of the system, or it can be used to study the properties of density operators itself. Furthermore, some parametrizations provide a straightforward way to generate positive matrices. When a general expression for a density matrix is needed,  the positivity
condition is typically the most difficult property to be sure of. Some parametrizations provide a simple way to overcome this problem.

Maybe the best known density matrix parametrization is the Bloch vector parametrization.
It was first used to describe the states of a two-level system, but was later extended to higher dimensions. An early review article discussing the properties of the Bloch vector was published by U. Fano in 1957 \cite{Fano57}. In the standard approach using hermitian basis matrices,
the Bloch vector of an $n$-level system is a real vector with $n^2-1$ components.
In a two-level setting, the Bloch vector is a three component vector with length smaller or equal to one.
It provides a way to identify the states of a two-level system with the points of a ball of radius one,
the pure states corresponding to the surface of the ball. This mapping has found applications in hundreds, if not thousands, of articles. The two-level case will be mentioned in this review only very briefly as the main emphasis is on higher-dimensional extensions of the Bloch vector description.
The structure of the set of Bloch vectors corresponding to positive operators becomes very complicated as soon as the dimension increases from two to three. This is a consequence of the fact that when $n>2$ the maximal length of the Bloch vector producing a positive operator depends on the direction of the vector. The complicated nature of this set is visible already in the three-level case \cite{Harriman78a,Ki03,BK03}.
However, despite the problems associated with the structure of the set of Bloch vectors giving a physical state,
the higher dimensional Bloch vector description has found applications in many fields, such as  the dynamics of $n$-level systems. It has been shown that in this context the Bloch vector parametrization helps to identify constants of motions of some quantum optical systems  \cite{Elgin80,Hioe81,Gottlieb82,Hioe82,Hioe85}. Assuming that the Hamiltonian has a specific shape,
the time evolution equation for the Bloch vector can be written in a block form.
Vector components belonging to different blocks evolve independently of each other,
and it turns out that the length of the vector inside each block is time independent  \cite{Hioe81,Hioe82,Hioe85}.
There exists also a class of constants of motion known as trace invariants. Unlike the constants of motion mentioned above, trace invariants are independent of the Hamiltonian of the system. The values of these invariants can be related to the components of the Bloch vector \cite{BK03}.
Another field where the Bloch vector parametrization has been used during the recent years is that of  quantum entanglement. In particular, the entangled states of two qubits have been studied using this approach by various authors, see, for example,
\cite{Fano83,Schlienz95,Aravind96,James01,Abouraddy02,Jaeger03,BK03,Theodorescu03,Altafini04,Jakob07}.
In the aforementioned applications
the Bloch vector is defined in a basis consisting of hermitian operators.
This is often a natural choice as it guarantees that the components of the Bloch vector are real.  However, it may be preferable to use the polarization  (or spherical tensor) operator basis \cite{Kryszewski06,Bertlmann08} when the angular symmetries of states are important \cite{Blum96}. Another possible choice for the basis is the Weyl operator basis \cite{Bertlmann08}. As in the case of the hermitian basis, defining the parameter set corresponding to physical states is a complicated problem also when the latter two bases are used.

It is possible to express every $n\times n$ density matrix in terms of a diagonal matrix and an
element of the unitary group $U(n)$. This provides  a way to parametrize the density matrices if a parametrization
 of $U(n)$ is known. There exists many such parametrizations, but, depending on the on the problem, one of them may be preferable by providing a better insight or by leading to a reduction of the number of parameters. In general, a parametrization of $U(n)$ has $n^2$
 real parameters. Consequently, the size of the parameter set grows fast as  $n$ increases, encouraging to find ways to reduce the number of parameters.
Two examples of parametrizations which allow to identify redundant parameters are the Jarlskog parametrization \cite{Ja05,FFK05,Ja06} and the parametrization presented in \cite{Spengler10}.
These parametrizations have potential applications in quantum information theory, where many quantities and properties of quantum systems are obtained by optimizing the values of some functions over the set of all density matrices. This is the case, for example, when  entanglement is quantified \cite{Horodecki09,Guhne09}.
Determining the values of entanglement measures  leads often to numerically demanding calculations due to the high-dimensional parameter set over which the optimization has to be performed. Finding a parametrization which allows to construct a sufficiently large set of density operators while keeping the parameter set small would simplify his task considerably.
The parametrizations discussed in this review are the coset \cite{Ak07} and  Jarlskog parametrizations \cite{Ja05,FFK05,Ja06}.
The former consists of parametrizing the density matrices in terms of certain cosets in the group $U(n)$, while in the latter approach the elements of $U(N)$ are given recursively, meaning that the elements
of $U(N)$ are expressed in terms of the elements of $U(N-1)$ and a unitary matrix containing
the additional parameters needed to describe an element of $U(N)$.
With the help of the Jarlskog parametrization it is straightforward to generate matrices which are guaranteed to be positive. Furthermore, this parametrization can be easily extended to composite systems \cite{BCP08}.

In addition to these approaches, there is a parametrization of (special) unitary matrices in terms of generalized Euler angles \cite{TS02,TS04}. It has been applied in \cite{TS04} to derive a volume formula for $U(N)$ and related groups. This parametrization allows to eliminate redundant global phases in several cases.
Yet another way to parametrize
the elements of $U(N)$ is in terms of quantum Householder reflections \cite{Ivanov06}.
This approach has been shown to be useful in quantum computation and quantum state manipulation \cite{Ivanov06,Ivanov07,Ivanov08}.
It seems, however, that these two parametrizations of the unitary group have not been used to parametrize density matrices.
Consequently they will not be discussed in more detail in this article.
Another related factorization of unitary matrices in terms of orthogonal matrices is described in \cite{Di03}. A substantial generalization of these ideas can be found in \cite{Di05}.

\subsection{Basic definitions}
Recall that in quantum physics states are represented by density matrices on the complex Hilbert space $\h$ of the system. The Hilbert space $\h$ is always separable and in many cases of modern applications actually finite dimensional.

A {\em density matrix} $\rho$ on a Hilbert space $\h$ is by definition a linear operator from $\h$ into $\h$ such that
\beq \label{d-matrix-pos}
\rho \geq 0\; \textrm{i.e.,}\quad \lan x, \rho x \ran \geq 0 \quad \textrm{for all}
\quad x \in \h
\eeq
and
\beq \label{d-matrix-normalization}
\Tr(\rho)= \sum_{j=1}^{\infty} \lan e_j,\rho e_j\ran = 1,
\eeq
where $\lan \cdot,\cdot \ran$ denotes the inner product of the Hilbert space $\h$,
$\{e_j:\, j \in \N\}$ is any orthonormal basis of $\h$ and the normalization condition
for the trace is written for the infinite dimensional case.
In this case one has also to assume that $\rho$ is of trace class.
If $\h$ is of finite dimension $n$, then the sum in (\ref{d-matrix-normalization})
extends from $j=1$ to $j=n$.

In this article we consider the case where the Hilbert space $\h$ is $n$-dimensional, $n \in \N$,
\beq \label{H-space}
\h = \h_n = \C^n\, .
\eeq
We denote the basis of $\C^n$ by $\{|1\rangle,|2\rangle,\ldots,|n\rangle\}$.
Now linear operators $A$ on $\h$ are represented by $n\times n$ matrices with complex entries,
and in the following we do not distinguish between a matrix and the linear operator it represents.
We denote by $\M_n$ the space of all $n\times n$ matrices with complex entries. Then a density matrix
$\rho$ on $\h_n$ is an element $\rho \in \M_n$ such that the positivity condition (\ref{d-matrix-pos})
and the normalization condition (\ref{d-matrix-normalization}) for the trace hold.

It is well known (and straightforward to prove) that in a complex Hilbert space a linear operator $A$
which is positive in the sense of (\ref{d-matrix-pos}) (often also called positive semi-definite)
is Hermitian, i.e., it satisfies $A=A^*$, where $A^*$ is the adjoint matrix defined by
\beq \label{adjoint}
(A^*)_{ij}= \bar{a}_{ji}\quad \textrm{for all}\quad i,j=1,\ldots,n
\eeq
Here $a_{ij}$ are the coefficients of the matrix $A$ and $\bar{a}_{ji}$ denotes the complex conjugate of $a_{ji}$.

The eigenvalues of a matrix $A \geq 0$ (i.e., $A$ satisfies (\ref{d-matrix-pos})) are calculated as follows
(see, for instance, Theorem 25.1.1 of \cite{BB03}): The first eigenvalue is given by
\beq \label{ev1}
\lambda_1 = \sup \{\lan x, Ax \ran: x \in \h_n,  \norm{x}=1 \},
\eeq
and one proves easily from this definition that there is a vector $e_1 \in \h_n$, $\norm{e_1}=1$ such
that $Ae_1 = \lambda_1 e_1$. The second eigenvalue then is ($\{e_1\}^{\perp}$ denotes the orthogonal complement of $\{e_1\}$ in $\h_n$)
\beq \label{ev2}
\lambda_2 = \sup \{\lan x, Ax \ran: x \in \{e_1\}^{\perp} \subset \h_n,  \norm{x}=1 \},
\eeq
 and  there is $e_2 \in \{e_1\}^{\perp}$, $\norm{e_2}=1$ such that $Ae_2 = \lambda_2 e_2$. This procedure can be iterated and produces eigenvalues
 \beq \label{ev-s}
 \lambda_1, \ldots, \lambda_n; \qquad \lambda_1\geq \lambda_2 \geq \ldots \geq \lambda_n \geq 0.
 \eeq
 These $n$ eigenvalues are not necessarily distinct; they occur in this list as many times as their multiplicity
 requires. For the eigenvalues of a  matrix $A \geq 0$ with $\Tr(A)=1$ we know in addition to (\ref{ev-s}) that
 \beq \label{ev-dens}
 \lambda_1 >0 \quad \rm{and} \quad \sum_{j=1}^n \lambda_j =1.
 \eeq
We know that the system of eigenvectors $e_1, \ldots, e_n$ of $A$ is a complete orthogonal system in $\h_n$.
Therefore, the transition from the standard basis of $\h_n$ to the basis of eigenvectors is effected by a
unitary $n \times n$ matrix $U$, and we arrive at the {\it spectral representation} of the matrix $A$:
 \beq \label{spec-rep}
 A = U^* D_n(\lambda_1, \ldots, \lambda_n) U,  \eeq
 with
 \beq \label{diag-ev}
 D_n(\lambda_1, \ldots, \lambda_n)= \begin{pmatrix}
\lambda_1 & 0 & 0 &\cdots &0\\0&\lambda_2 & 0 & \cdots & 0\\
\vdots & \vdots & \vdots & \lambda_{n-1} &0\\
0 & \cdots & 0 & 0&\lambda_n
\end{pmatrix}   \eeq
being the diagonal matrix of eigenvalues.

Thus  a density matrix can be characterized as follows:
\begin{quote}
A matrix $\rho \in \M_n^h$ ($\M_n^h$ denotes the space of all Hermitian elements of $\M_n$) with coefficients $\rho_{ij} \in \C$, $i,j=1,\ldots,n$ is a {\em density matrix} if, and only if,
\begin{enumerate}
 \item[(a)] $\rho$ is positive in the sense of (\ref{d-matrix-pos}), i.e. all eigenvalues $\lambda_j$ are nonnegative,
  \beq \label{pos-eigval}
  \lambda_j =\lambda_j(\rho)\geq 0 \quad \textrm{for} \quad j=1,\ldots,n; \eeq
 \item[(b)] \beq \label{trace-one}
 \Tr(\rho)=\sum_{j=1}^n \rho_{jj} =1. \eeq
\end{enumerate}
\end{quote}
Let us introduce the set $\cald_n$ of all $n$-dimensional density matrices,
\beq \label{density-matrix}
\cald_n = \{\rho \in \M_n^h :\, \rho \; \textrm{satisfies}\; (\ref{pos-eigval}) \; \textrm{and} \; (\ref{trace-one}) \}.
\eeq
In this article we study the general form of  the elements of $\cald_n$. The main difficulty is the positivity
constraint $\rho \geq 0$. There are a number of criteria a matrix has to satisfy for positivity.
Since they are typically given in terms of inequalities they do not provide much information about the concrete form of a density matrix.

In this article we review recent studies on the structure and general form of a density matrix. Ideally one would like to have a parametrization of density matrices in the following sense: A {\em parametrization of a density matrix} $\rho \in \cald_n$ means the following:
\begin{enumerate}
\item[(a)] Specification of a parameter set $Q_n \subset \R^m$ where $m$ depends on $n$, i.e., $m=m(n)$;
\item[(b)] Specification of a one-to-one and onto map $F_n:\, Q_n \To \cald_n$.
\end{enumerate}
Clearly, the case $n=1$ is trivial: $\cald_1 = \{1\}$. Therefore in the following we assume $n\geq 2$.

Unfortunately, this type of a parametrization of density matrices is not yet available for general $n \geq 3$.
Only the case $n=2$ is fully understood. We discuss here the partial results of the three main approaches to this problem:
\begin{itemize}
 \item the Bloch-vector parametrization;
 \item the coset parametrization;
 \item the Jarlskog parametrization.
\end{itemize}

\subsection{Overview}
Although the case $n=2$ has been studied a lot in the literature, we start the following section on the
Bloch-vector parametrization with a review of this case. This serves as a preparation for the discussion on the general case $n\geq 3$.
This discussion  is based mostly on \cite{Ki03,BK03} and it provides a reference point for the
representations presented in sections \ref{sec:Coset} and  \ref{sec:Jarlskog}, where it turns out that all parametrizations agree for $n=2$, but not for $n>2$.
 We show that, despite  the fact that the determination of the parameter set $Q_n$ is a difficult task, the  Bloch-vector parametrization can be used to identify various constants of motion or to study the properties of two-qubit states.

Obviously, the spectral representation (\ref{spec-rep}) can be used to obtain a parametrization of density matrices as soon as a parametrization of unitary matrices is known. Several parametrizations of unitary matrices can be found in the literature.
We mention those which are  of interest in connection with our problem of parametrizing density matrices.
One of them is the coset parametrization (see, for instance, \cite{Gi74}). It has been used
 in \cite{ACH93} as a starting point for an analysis of the space of all density matrices for $n=4$.
This analysis gives a description of the geometry of this state space in terms of flag manifolds.
However, no parametrization of individual density matrices is offered.
This idea has later been used in \cite{Ak07} to propose a parametrization of density matrices in terms of certain cosets in the group $U(n)$.
Section \ref{sec:Coset} contains a brief summary and discussion about the extent to which this parametrization can be considered a parametrization in the sense given above.

The next section introduces the Jarlskog parametrization (see \cite{BP08}). This parametrization is closely related to the coset parametrization and again starts from the spectral representation (\ref{spec-rep}) but uses the parametrization of  $U(n)$ presented by Jarlskog (see \cite{Ja05,Ja06}) instead of cosets. The Jarlskog parametrization is recursive and thus allows a recursive parametrization of density matrices. We apply the Jarlskog parametrization to composite systems and obtain several quite interesting cases of a parametrization of density matrices for composite systems \cite{BCP08}. Finally, section \ref{sec:Conclusion} contains the concluding remarks.

\section{Bloch-vector parametrization}\label{sec:Bloch}
In this section we first describe the Bloch vector of a two-level system,
and then extend the discussion to $n$-dimensional systems.
We show what is required from the Bloch vector of a three-level system for it to
desribe a physical state. We illustrate the use of the Bloch vector in identifying constants of motion of dynamical systems, and then show how the Bloch vector approach can be used to study  the separability of a two-qubit system. We also briefly describe an alternative basis for the Bloch vector parametrization, known as the polarization operator basis.

\subsection{The case $n=2$: The Bloch/Poincar\'e sphere}
It is not difficult to see that the Pauli matrices
\beq \label{P-matrices}
\sigma_x = \sigma_1=\begin{pmatrix} 0& 1\\1&0 \end{pmatrix}, \quad \sigma_y =\sigma_2 =
\begin{pmatrix} 0& -\ii\\ \ii & 0 \end{pmatrix}, \quad \sigma_z =\sigma_3 =
\begin{pmatrix} 1 &0\\0&-1 \end{pmatrix},
\eeq
together with the identity matrix $I_2$ form a basis of the real vector space $\M_2^h$ of Hermitian $2 \times 2$ matrices.
Hence every $A \in \M_2^h$ can be represented as
$$A= a_0 I_2 + \lambda_1 \sigma_1 + \lambda_2 \sigma_2 + \lambda_3 \sigma_3,$$
where $a_0$ and $\lambda_j$, $j=1,2,3$ are real numbers. Since the trace of the Pauli matrices vanishes,
the coefficient $a_0$ is determined by $\Tr(A)=2 a_0$. Hence, such a matrix belongs to $\cald_2$  if and only if
it has a representation
\beq \label{2d-density-m}
A = \frac{1}{2} I_2 +\frac{1}{2}\sum_{j=1}^3 \lambda_j \sigma_j = \frac{1}{2} \begin{pmatrix}
1+\lambda_3 & \lambda_1 - \ii \lambda_2 \\ \lambda_1 + \ii \lambda_2 & 1- \lambda_3
\end{pmatrix}, \eeq
in which the coefficients $\lambda_j$ are chosen such that all the eigenvalues of $A$
are non-negative. The eigenvalues of $A$ can be calculated by finding the roots of the characteristic polynomial $\det(xI_2 -A)$. Using the abbreviation $\lambda_j\pr = \frac{1}{2} \lambda_j$ we find for this
polynomial ($\und{\lambda}=(\lambda_1,\lambda_2,\lambda_3) \in \R^3$ and $|\und{\lambda}|^2= \sum_{j=1}^3 \lambda_j^2$)
\beq \label{char-pol-2d}
\det(xI_2 -A) =(x-\frac{1}{2})^2 -|\und{\lambda\pr}|^2 =x^2 -x + \frac{1}{4} -|\und{\lambda\pr}|^2.\eeq
The roots of this polynomial are
\beq \label{roots-2d}
x_1 = \frac{1}{2}(1 + |\lambda|) \quad \textrm{and} \quad x_2 = \frac{1}{2}(1 - |\lambda|). \eeq
While the root $x_1 $ is always $\geq \frac{1}{2}$, the root $x_2$ is non-negative if and only if $|\und{\lambda}| \leq 1$.
This gives our first result: The parameter set for this case is
\beq \label{Q-2}
Q_2 = \{\und{\lambda} \in \R^3 : \, |\und{\lambda}| \leq 1 \} =B(\R^3), \eeq
where $B(\R^3)$ denotes the closed unit ball in $\R^3$ with center at $0$. We define a map on $Q_2$ with values
in $\cald_2$ by the right hand side of (\ref{2d-density-m}), i.e.,
\beq \label{d2-map}
F_2(\und{\lambda})= \frac{1}{2} \begin{pmatrix}
1+\lambda_3 & \lambda_1 - \ii \lambda_2 \\ \lambda_1 + \ii \lambda_2 & 1- \lambda_3
\end{pmatrix},\quad \und{\lambda} \in Q_2. \eeq
By construction, this map is onto $\cald_2$. A simple argument shows that $F_2$ is also one-to-one, i.e.,
$F_2(\und{\lambda})=F_2(\und{\lambda}\pr)$ implies $\und{\lambda}=\und{\lambda}\pr$. We conclude that that
$(Q_2, F_2)$ is a parametrization of $\cald_2$ with $m=3 = n^2 -1$ for $n=2$.

The parameter set $Q_2= B(\R^3)$ in this parametrization of $\cald_2$ is called the {\em Bloch} or {\em
Poincar\'e ball}. In order to relate it to the standard representation and interpretation we just have to introduce spherical polar coordinates in $B(\R^3)$:
\beq \label{spherical-coordinates}
\lambda_1 = r \sin \theta \cos \phi,\quad \lambda_2 = r \sin \theta \sin \phi,\quad\lambda_3 = r \cos \theta ,\quad 0\leq \theta \leq\pi,\;0\leq \phi < 2\pi,\;0\leq r \leq 1.
\eeq
Using the spherical coordinates we can introduce a new parameter set
\beq \label{Q-2-2}
\tilde{Q}_2 =\{(\theta,\phi,r) \in \R^3:\,0\leq \theta \leq\pi,\;0\leq \phi < 2\pi,\;0\leq r \leq 1\} \eeq
and on it the map $\tilde{F}_2$ with values in $\cald_2$, defined by
\beq \label{d2-map2}
\tilde{F}_2(\theta,\phi,r)=F_2(\und{\lambda}(\theta,\phi,r)) = \begin{pmatrix}
\frac{1}{2}(1 +r \cos \theta) & \frac{r}{2} e^{- \ii\phi}\sin\theta \\ \frac{r}{2} e^{ \ii\phi}\sin\theta &
\frac{1}{2}(1 -r \cos \theta)
\end{pmatrix}, \quad (\theta,\phi,r) \in \tilde{Q}_2.
\eeq
The boundary of the Bloch/Poincar\'e ball is called the {\em Bloch/Poincar\'e sphere}.
In terms of the parameter sets $Q_2$ and $\tilde{Q}_2$ this sphere is given by $|\und{\lambda}| =1$ and $r=1$, respectively.
Introducing the angle $\vartheta= \theta/2$ and using some elementary relations for the trigonometric functions $\sin$ and $\cos$,
the pure state density matrix corresponding to the point $(\theta,\phi,1)$ on the Bloch/Poincar\'e sphere has the form
\beq \label{d2map-sphere}
\tilde{F}_2(2\vartheta,\phi,1)= \begin{pmatrix}
c^2 &cs e^{-\ii \phi}\\ cs e^{\ii \phi} & s^2
\end{pmatrix} =|\psi\rangle\langle\psi| , \eeq
where we used the abbreviations $c= \cos \vartheta$ and $s=\sin \vartheta$, and
\beq \label{pure-state}
|\psi\rangle = (c, s e^{\ii\phi})^T = \cos\vartheta |0\rangle + e^{\ii \phi}\sin \vartheta |1\rangle.  \eeq

\subsection{The general case $n \geq 3$}
As the last subsection shows, the Bloch vector parametrization works
quite well in the case of 2-level systems ($n=2$). It has been
generalized to the case of $n$-level systems with $n\geq 3$. The
starting point of this generalization is the observation that the
Pauli-matrices in (\ref{2d-density-m}) can be considered as the
generators of  the special unitary group $SU(2)$. Therefore the
following representation of density matrices $\rho \in \cald_n$ has
been suggested: \beq \label{BV-rep-n} \rho = \frac{1}{n}I_n +
\frac{1}{2} \sum_{j=1}^{n^2 -1} \lambda_j \hat{\lambda}_j, \eeq
where $\hat{\lambda}_j$, $j=1,\ldots, n^2 -1$, are the (orthogonal)
generators of the special unitary group $SU(n)$ and the $\lambda_j$,
$j=1,\ldots,n^2-1$,  are real numbers. The generators
$\hat{\lambda}_j$ are $n\times n$ matrices with complex coefficients
satisfying \beq \label{sun-generator} \hat{\lambda}_j^{*} =
\hat{\lambda}_j, \quad \Tr \hat{\lambda}_j =0, \quad
\Tr(\hat{\lambda}_i\hat{\lambda}_j) =2 \delta_{ij}, \quad
i,j=1,\ldots, n^2 -1 \eeq and the commutation and anti-commutation
relations \beq \label{commut-rel}
[\hat{\lambda}_i,\hat{\lambda}_j]=2 \ii \sum_{k=1}^{n^2-1}
f_{ijk}\hat{\lambda}_k, \quad [\hat{\lambda}_i,\hat{\lambda}_j]_+=
\frac{4}{n}\delta_{ij} I_n +2  \sum_{k=1}^{n^2-1} g_{ijk}
\hat{\lambda}_k. \eeq Here the $f_{ijk}$ and $g_{ijk}$ are the
structure constants of the Lie algebra $su(n)$. It is also known
that the generators $\hat{\lambda}_i$, $i=1,\ldots,n^2-1$ together
with the unit matrix $I_n$ form an orthogonal basis of $\M_n^h$ with
respect to the Hilbert-Schmidt inner product $\lan A, B\ran =
\Tr(A^{*}B)$. This means that the inner product of two density
matrices $\rho$ and $\tilde{\rho}$, with Bloch vectors
$\underline{\lambda}$ and $\underline{\tilde{\lambda}}$, reads
simply
$\langle\rho,\tilde{\rho}\rangle=\underline{\lambda}\cdot\underline{\tilde{\lambda}}$

The vector $\underline{\lambda}$ is in this review called Bloch vector. In order to make a difference between the Bloch vector of a two-level system
and that of an $n$-level system, $n>2$, the term {\it generalized Bloch vector} is sometimes used in the literature to describe the latter case. Yet another possible name
for the Bloch vector is {\it coherence vector}.

One possible explicit construction for the generators
$\hat{\lambda}_j$ can be given in terms of the generalized Gell-Mann
matrices by defining the basis as in the following equation.
\begin{align}
\label{GGM}
\nonumber
&|j\rangle\langle k|+|k\rangle\langle j|,& 1\leq j<k\leq n,\\
\nonumber
-\ii&|j\rangle\langle k|+\ii|k\rangle\langle j|,& 1\leq j<k\leq n,\\
\sqrt{\frac{2}{l(l+1)}}&\left(\sum_{j=1}^l |j\rangle\langle j|-l|l+1\rangle\langle l+1|\right),& 1\leq l\leq n-1.
\end{align}
If $n=2$ these correspond to the Pauli matrices (\ref{P-matrices}) and if $n=3$ they are the Gell-Mann matrices, see
(\ref{Gell-Mann}) below.

According to the properties (\ref{sun-generator}) of the generators $\hat{\lambda}_j$ every matrix of the form (\ref{BV-rep-n}) has trace $1$, $\Tr \rho =1$. Thus a matrix of the form (\ref{BV-rep-n}) is a density matrix if and only if all its eigenvalues are non-negative. As in the case $n=2$ these eigenvalues are the roots of the characteristic polynomial $\det(x I_n - \rho)$. As a polynomial of degree $n$ the characteristic polynomial has a unique representation of the form
\beq \label{char-poly-n}
\det(x I_n - \rho)= \sum_{j=0}^n (-1)^j a_j \,x^{n-j}, \quad a_0=1,
\eeq
where the coefficients $a_j$ are uniquely determined by the generators $\hat{\lambda}_j$ and the parameters
$\und{\lambda}=(\lambda_1,\ldots, \lambda_{n^2-1}) \in \R^{n^2-1}$. In order to emphasize the dependence on
$\und{\lambda}$ we write $a_j=a_j(\und{\lambda})$.

If $x_1,\ldots,x_n$ denote the roots of the characteristic
polynomial $\det(x I_n - \rho)$, then (\ref{char-poly-n}) shows \beq
\label{char-pol-roots} \sum_{j=0}^n (-1)^j a_j \,x^{n-j}
=\prod_{j=1}^n (x-x_j) \eeq and by evaluating the product and
comparing coefficients, the basic relation between the coefficients
$a_j$ and the roots $x_j$ follows: \beq \label{Vieta-formula} a_j=
\sum_{1\leq i_1 <i_2< \cdots i_j}^n x_{i_1} x_{i_2}\cdots x_{i_j}
\qquad \textrm{Vieta's formula}. \eeq It is known (an elementary
proof is given in the appendix of \cite{Ki03}) that from this
formula the important characterization of the
non-negativity of the eigenvalues follows: \beq \label{ev-pos} x_j
\geq 0,\; j=1,\ldots,n, \quad \Leftrightarrow \quad a_j \geq
0,\;j=1,\ldots,n. \eeq Accordingly we define  the parameter set as
$Q_n$ \beq \label{par-set-n} Q_n =\{\und{\lambda} \in \R^{n^2-1}:\;
a_j(\und{\lambda}) \geq 0,\, j=1,\ldots, n\}, \eeq and on $Q_n$ the
map $F_n : \,Q_n \To \cald_n$ by \beq \label{par-map-n}
F_n(\und{\lambda})=\frac{1}{n}I_n +\frac{1}{2}\sum_{j=1}^{n^2-1}
\lambda_j \hat{\lambda}_j, \quad \und{\lambda} \in Q_n. \eeq The
inverse of this map is given by \beq \label{inverse-par-map}
F_n^{-1}(\rho) = (\lambda_1=\Tr(\rho \hat{\lambda}_1),
\ldots,\lambda_{n^2-1}=\Tr(\rho \hat{\lambda}_{n^2-1})), \quad \rho
\in \cald_n. \eeq Thus, $(Q_n,F_n)$ is a parametrization for
$n$-level density matrices $\rho \in \cald_n$. It is called the {\em
Bloch-vector parametrization}.

Now we discuss the difficulties with this parametrization for concrete applications. Although the parameter set $Q_n$ is uniquely specified, it is not easy to decide when a given point $\und{\lambda} \in \R^{n^2-1}$ actually belongs to $Q_n$. The origin of this difficulty is that firstly it is quite a complicated matter to explicitly calculate the coefficients $a_j$ for larger values of $j$ and secondly it is not easy to determine
explicitly the boundary of $Q_n$, i.e., the set $\partial Q_n = \{\und{\lambda} \in \R^{n^2-1} :\, a_j(\und{\lambda})=0, j=1,\ldots,n^2-1 \} $ since this amounts to solving polynomial equations in $\und{\lambda}$ of degree $n$, in $n^2-1$ variables.

Using Lie algebra techniques, the polynomials $a_j(\und{\lambda})$, $j=1,2,3,4$, have been calculated explicitly in \cite{Ki03} (see also \cite{BK03}). They read as follows:
\begin{align} \label{a1234}
1! a_1 &= 1, \nonumber \\
2! a_2 &= \left( \frac{n-1}{n} -\frac{1}{2}|\und{\lambda}|^2 \right), \nonumber \\
3! a_3 &= \left[ \frac{(n-1)(n-2)}{n^2} - \frac{3(n-2)}{2n}|\und{\lambda}|^2 + \frac{1}{2} \sum_{i,j,k=1}^{n^2-1}
         g_{ijk}\lambda_i \lambda_j \lambda_k \right], \nonumber \\
4! a_4 &=\left[ \frac{(n-1)(n-2)(n-3)}{n^3} - \frac{3(n-2)(n-3)}{n^2} |\und{\lambda}|^2
    + \frac{3(n-2)}{4n}|\und{\lambda}|^4 \right . \nonumber \\
      &  \left. \quad
      + \frac{2(n-3)}{n} \sum_{i,j,k=1}^{n^2-1}  g_{ijk}\lambda_i \lambda_j \lambda_k
- \frac{3}{4}\sum_{i,j,k,l,p=1}^{n^2-1} g_{ijk}g_{klp}\lambda_i \lambda_j \lambda_l  \lambda_p \right ] .
\end{align}

Already for $n=3$ the parameter set has not been determined explicitly though $a_3(\und{\lambda})$ is known.
If in (\ref{a1234}) we insert the explicit values of the structure constants $g_{ijk}$ for the chosen set of generators of $su(3)$, $a_3$ is given by (recall that $n^2 -1$ is $8$ for $n=3$)
\begin{multline}\label{a3}
3! a_3(\und{\lambda}) =\frac{1}{36}\left\{ 8 - 18 |\und{\lambda}|^2 + 27\lambda_3(\lambda_4^2 + \lambda_5^2 - \lambda_6^2 -\lambda_7^2) -6 \sqrt{3}\lambda_8^3 \right. \\
\left. +9\sqrt{3}\lambda_8 [2(\lambda_1^2 +\lambda_2^2 +\lambda_3^2)-(\lambda_4^2 + \lambda_5^2 +\lambda_6^2 +\lambda_7^2)]
 +54(\lambda_1\lambda_4\lambda_6 + \lambda_1\lambda_5\lambda_7 +\lambda_2\lambda_5\lambda_6 -\lambda_2\lambda_4\lambda_7)\right\}
\end{multline}
Thus it is obvious that the set $a_3(\und{\lambda})\geq 0$ and with it the parameter set $Q_3$ has not been determined explicitly. However in \cite{Ki03} some 2-dimensional sections have been calculated and represented graphically\footnote[1]{We remark that equation (31) of \cite{Ki03} contains typing errors.
The correct expression is given by (\ref{a3}) above.}. They show the very complicated nature of these sets.

The Bloch vector parametrization has found applications in many different fields:
It was used extensively by Harriman to study the properties of density operators \cite{Harriman78a,Harriman78b,Harriman83,Harriman86}.
In particular, he discussed the structure of the set of physical states $Q_n$ \cite{Harriman78a}. This work was
later extended, and partly reproduced, by other authors \cite{Ki03,BK03}.
It has been useful in studying the dynamics and constants of motions of $n$-level systems \cite{Elgin80,Hioe81,Hioe82,Gottlieb82,Hioe84,Hioe85,Oreg85},
Markovian dynamics of decaying two-level systems \cite{Pottinger85,Lendi86b,Alicki}, entropy production \cite{Lendi86},
characterization of the reachable sets for open systems driven by unitary control \cite{Altafini03}, and
unitary orbits in the set of density operators \cite{Schirmer04}. Furthermore,
it has been used in studying the properties of two- and  $n$- qubit systems
\cite{Fano83,Schlienz95,Aravind96,James01,Abouraddy02,Jaeger03,BK03,Theodorescu03,Altafini04},
the geometry of the states of a finite-dimensional systems \cite{BD08},
the classification of density matrices into different types \cite{Mendas06,Mendas08},
detecting the entanglement properties of bipartite quantum states \cite{Krammer09},
characterizing the structure of the state space of a two-qubit system \cite{Jakobczyk01} and
the steady states of open quantum systems \cite{Schirmer10}.

In the following we  briefly discuss some of these applications.

\subsection{Bloch parametrization and trace invariants}
As we show next, $n$-level systems can be associated with constants of motion which are independent
of the form of the Hamiltonian and depend only on the initial state of the system.
These constants of motion are obtained by tracing over $\rho^n$, $n=1,2,\ldots$,
and are therefore called trace invariants. These invariants can be expressed using the Bloch vector
and their values are also related to the coefficients $a_i$ of
the polynomial (\ref{char-pol-roots}) \cite{Ki03,BK03}. The following discussion is based on \cite{Hioe81,Ki03,BK03}.

Before introducing the trace invariants, we derive the differential equation that determines the time-evolution of the Bloch vector.
The time-evolution of an $n$-level system can be expressed in terms of its density matrix
$\rho$ which satisfies the Liouville equation
\beq\label{Liouville}
\ii\frac{d\rho(t)}{d t}=[\hat{H}(t),\rho(t)],
\eeq
where the Hamiltonian $\hat{H}$ is in general time dependent and we have set $\hbar=1$.
It has been shown in \cite{Reed} (see also \cite{DGT88,KN08}) that under certain conditions for the
Hamiltonian\footnote[2]{{\color{black}$H(t)$ is
bounded and self-adjoint and the map $t\mapsto H(t)$ is strongly continuous}} this equation can be solved in the form
\beq \label{time-dep-Ham-sol}
\rho(t)=U_t \rho(0)U_t^*
\eeq
where $U_t$ , $t \in \R$, is a family of unitary operators (which in general is neither a group nor a semi-group) and where $\rho(0)$ is the initially given density matrix at time $t=0$. This form of the time evolution guarantees that one stays in the space of density matrices.
We write $\rho$ as in (\ref{BV-rep-n})
\beq \label{BV-rep-n2}
\rho = \frac{1}{n}I_n + \frac{1}{2} \sum_{j=1}^{n^2 -1} \lambda_j \hat{\lambda}_j.
\eeq
Here and in what follows the time-dependence of the states and Hamiltonians is not always explicitly indicated.
The Hamiltonian $\hat{H}$ can be expressed in a similar way
\beq\label{Ham}
\hat{H}=\frac{h_0}{n}I_n+\frac{1}{2}\sum_{i=1}^{n^2-1}h_j\hat{\lambda}_{j},
\eeq
where
\beq
h_0=\Tr\hat{H},\quad h_j=\Tr(\hat{\lambda}_j\hat{H}),\quad j=1,\ldots , n^2-1.
\eeq
Using (\ref{BV-rep-n2}), (\ref{Ham}) and  the Liouville equation,
 the following equation for the time-evolution of the Bloch vector can be obtained
\beq\label{lambdadot}
\frac{d\lambda_i}{dt}=\sum_{j,k=1}^n f_{ijk} h_j\lambda_k.
\eeq
The structure constant $f_{ijk}$ is antisymmetric, which guarantees that the length of $\underline{\lambda}$
is time independent. In addition to this constant of motion, also other conserved quantities can be easily identified.
 We denote the rank of $\rho$ by $r$. Then the density matrix can be associated with $r$ constants of motion,
as can be seen by writing the density matrix at instant $t$ as
\beq\label{UDU}
\rho(t)=V(t)DV^*(t), \quad D= D_n(\mu_1,\mu_2,\ldots,\mu_n),
\eeq
where $D$ is a diagonal matrix of the eigenvalues of $\rho$ and $V$ is a map $V:\mathbb{R}\rightarrow U(n)$.
The eigenvalues are time independent and $V$ gives the time-evolution of the state. Now
\beq
\Tr(\rho^k(t) )=\Tr(D^k)=\mu_1^k+\mu_2^k+\cdots+\mu_n^k
\eeq
defines a constant of motion for each integer $k>0$.
The number of non-zero eigenvalues is equal to the rank  of $\rho$, so there are
$r$ independent constants of motion, given by $\Tr(\rho^k)$, $k=1,2,\ldots,r$.
Due to the way they are defined, these constants are called trace invariants.
Instead of using (\ref{UDU}), it is possible to prove the existence of the trace invariants
using (\ref{lambdadot}) and the properties of the generators $\{\hat{\lambda_j}\}$ of $SU(n)$.
The values of the trace invariants can be expressed in terms of the Bloch vector $\underline{\lambda}$.  For $k=1,2,3,4$ they are
\begin{align}\label{rho1234}
\nonumber
\Tr\rho &=1\\
\nonumber
\Tr(\rho^2)&=\frac{1}{n}+\frac{1}{2}|\underline{\lambda}|^2\\
\nonumber
\Tr(\rho^3)&=\frac{1}{n^2}+\frac{3}{2n}|\underline{\lambda}|^2
+\frac{1}{4}\underline{\lambda}\cdot(\underline{\lambda}\odot \underline{\lambda})\\
\Tr(\rho^4)&=\frac{1}{n^3}+\frac{3}{n^2}|\underline{\lambda}|^2
+\frac{1}{n}\underline{\lambda}\cdot(\underline{\lambda}\odot \underline{\lambda})
+\frac{1}{4n}|\underline{\lambda}|^4 +\frac{1}{8}|\underline{\lambda}\odot \underline{\lambda}|^2,
\end{align}
where we have defined
\beq
(\underline{a}\odot\underline{b})_k=\sum_{i,j=1}^{n^2-1}g_{ijk}a_i b_j.
\eeq
Explicit formulas for the trace invariants for $n=1,\ldots ,9$ given in terms of the Bloch vector  can be found
in the Appendix B of \cite{BK03}. Conditions for a vector $\underline{\lambda}$ to describe
a pure state can be obtained from these equations by setting $\Tr(\rho^k)=1$. In this way one also obtains
the boundary $\partial Q_n$ of $Q_n$, which consists of pure states.

Comparing (\ref{a1234}) and (\ref{rho1234}) we see that the coefficients $a_i$ can be given
in terms of the trace invariants. The expressions for the first four coefficients and the general expression
are
\begin{align}
\nonumber
1!a_1&=\Tr\rho=1\\
\nonumber
2!a_2&=1-\Tr(\rho^2)\\
\nonumber
3!a_3&=1-3\Tr (\rho^2)+2\Tr(\rho^3)\\
\nonumber
4!a_4&=1-6\Tr(\rho^2)+8\Tr(\rho^3)-6\Tr(\rho^4)+3\Tr(\rho^2)^2\\
k!a_k&=(k-1)!\left((-1)^{k-1}\Tr(\rho^k)+\sum_{i=1}^{k-1}(-1)^{i-1}\Tr(\rho^i)a_{k-i}\right).
\end{align}
Clearly also the coefficients $a_i$ are time independent. With the help of (\ref{rho1234}) these can be
expressed using the Bloch vector.

\subsection{Bloch parametrization and the dynamics of three-level systems}
We now describe how the Bloch vector parametrization can be used to study
the dynamics of three-level systems.
This topic was studied long before the structure of the set $Q_n$, that is, the structure of the set
of Bloch vectors corresponding to physical states, was examined.
 We remark that in these studies it was not necessary
to know the exact shape of the set  $Q_n$. This is a consequence of the unitary time-evolution,
which guarantees that the Bloch vector $\underline{\lambda}(t)$ belongs to the set $Q_n$ at all times $t>0$
if it is an element of $Q_n$ at $t=0$. One has only to make sure that
the initial state determined by $\underline{\lambda}(0)$ is a positive operator.

Here and in what follows the initial time is chosen to be $t=0$.
The following discussion is based on \cite{Elgin80,Hioe81,Hioe82,Gottlieb82,Hioe84,Hioe85}.
It was found in \cite{Elgin80,Hioe81,Gottlieb82} that the Bloch vector representation of the density matrices allows to
identify, in addition to the trace invariants whose values do not depend on the shape of the Hamiltonian,
also additional conserved quantities which exist only when the Hamiltonian has a specific form.
We illustrate this by studying a three-level system,
so the relevant Lie algebra is that of $SU(3)$. One common realization for the
basis elements $\{\hat{\lambda}_1,\ldots,\hat{\lambda}_8\}$ of this Lie algebra
 is given by the Gell-Mann matrices:
\begin{align}
\label{Gell-Mann}
\nonumber
&\hat{\lambda}_1=
\begin{pmatrix}
0 & 1 & 0\\
1 & 0 & 0\\
0 & 0 & 0
\end{pmatrix},\quad
\hat{\lambda}_2=
\begin{pmatrix}
0 & 0 & 1\\
0 & 0 & 0\\
1 & 0 & 0
\end{pmatrix},\quad
\hat{\lambda}_3=
\begin{pmatrix}
0 & 0 & 0\\
0 & 0 & 1\\
0 & 1 & 0
\end{pmatrix}\\
\nonumber
&\hat{\lambda}_4=
\begin{pmatrix}
0 & -\ii & 0\\
\ii & 0 & 0\\
0 & 0 & 0
\end{pmatrix},\quad
\hat{\lambda}_5=
\begin{pmatrix}
0 & 0 & -\ii\\
0 & 0 & 0\\
\ii & 0 & 0
\end{pmatrix},\quad
\hat{\lambda}_6=
\begin{pmatrix}
0 & 0 & 0\\
0 & 0 & -\ii\\
0 & \ii & 0
\end{pmatrix}\\
&\hat{\lambda}_7=
\begin{pmatrix}
1 & 0 & 0\\
0 & -1 & 0\\
0 & 0 & 0
\end{pmatrix},\quad
\hat{\lambda}_8=
\frac{1}{\sqrt{3}}
\begin{pmatrix}
1 & 0 & 0\\
0 & 1 & 0\\
0 & 0 & -2
\end{pmatrix}.
\end{align}

The physical system we are interested in consists of a three-level atom interacting with two lasers.
The energies of the three atomic levels are $\omega_1,\omega_2$ and $\omega_3$ and
the energy differences needed now are denoted by $\omega_{12}=\omega_1-\omega_2$ and $\omega_{23}=\omega_2-\omega_3$.
The frequencies of the two lasers are $\nu_1$ and $\nu_2$ and  $\Delta_{12}=\nu_1-\omega_{12}$,
$\Delta_{23}=\nu_2-\omega_{23}$ are the detunings. We assume the case of exact two-photon resonance,
which means that
\beq
\Delta_{12}=-\Delta_{23}=\Delta.
\eeq
Under these assumptions the Hamiltonian becomes
\begin{align}\label{HHioe}
\hat{H}&=
-\begin{pmatrix}
0 & \frac{1}{2}\Omega_{12} & 0\\
\frac{1}{2}\Omega_{12} & \Delta & \frac{1}{2}\Omega_{23}\\
0 & \frac{1}{2}\Omega_{23} & 0
\end{pmatrix}.
\end{align}
Here $\Omega_{12}=\underline{d}_{12}\cdot\underline{\mathcal{E}_{1}}$ and $\Omega_{23}=\underline{d}_{23}\cdot\underline{\mathcal{E}_{2}}$
are the Rabi frequencies, $\underline{d}_{jk}$ is the atomic dipole moment between levels $j$ and $k$ and
$\underline{\mathcal{E}_{1}}$ and $\underline{\mathcal{E}_{2}}$ are the vector amplitudes of electric fields of the lasers.
We assume that the Rabi frequencies $\Omega_{12},\Omega_{23}$ have a specific form given by
\begin{align}
\Omega_{12}(t)&=2a\,\Omega_0 (t)\\
\Omega_{23}(t)&=2b\,\Omega_0 (t),
\end{align}
where $a$ and $b$ are non-negative real numbers.
This means that $\Omega_{12}$ and $\Omega_{23}$ have the same time dependence
but possibly different amplitudes.
In the present case  equation (\ref{lambdadot}) gives
\beq\label{dldt}
\left(\frac{d\underline{\lambda}}{dt}\right)^T=V\underline{\lambda}^T,
\eeq
where
\beq
V=
\begin{pmatrix}
0 & 0 & 0 & -\Delta & -b\Omega_0 & 0 & 0 & 0\\
0 & 0 & 0 & -b\Omega_0 & 0 & a\Omega_0 & 0 & 0\\
0 & 0 & 0 & 0 & a\Omega_0 & \Delta & 0 & 0\\
\Delta & b\Omega_0 & 0 & 0 & 0 & 0 & 2a\Omega_0 & 0\\
b\Omega_0 & 0 & -a\Omega_0 & 0 & 0 & 0 & 0 & 0\\
0 & -a\Omega_0 & -\Delta & 0 & 0 & 0 & -b\Omega_0 & \sqrt{3}b\Omega_0\\
0 & 0 & 0 & -2a\Omega_0 & 0 & b\Omega_0 & 0 & 0\\
0 & 0 & 0 & 0 & 0 & -\sqrt{3}b\Omega_0 & 0 & 0
\end{pmatrix}
\eeq
and $T$ denotes transpose. In order to simplify (\ref{dldt}) we define a new basis $\{\hat{\lambda}_1',\ldots ,\hat{\lambda}_8'\}$ through the equation
\beq
\underline{\hat{\lambda}}^{\prime T}=B\underline{\hat{\lambda}}^T.
\eeq
Here $\underline{\hat{\lambda}}=(\hat{\lambda}_1,\ldots,\hat{\lambda}_8)$ and $\underline{\hat{\lambda}}'
=(\hat{\lambda}_1',\ldots,\hat{\lambda}_8')$ are vectors formed from the generators of $SU(3)$
and the basis change is given by the time independent orthogonal matrix
\beq
\label{B}
B=
\frac{1}{\sqrt{a^2+b^2}}\begin{pmatrix}
a & 0 & b & 0 & 0 & 0 & 0 & 0\\
0 & 0 & 0 & a & 0 & -b & 0 & 0\\
0 & \frac{ab}{\sqrt{a^2+b^2}} & 0 & 0 & 0 & 0 & \frac{2a^2+b^2}{2\sqrt{a^2+b^2}} & -\frac{\sqrt{3}b^2}{2\sqrt{a^2+b^2}}\\
0 & 0 & 0 & 0 & -1 & 0 & 0 & 0\\
b & 0 & -a & 0 & 0 & 0 & 0 & 0\\
0 & 0 & 0 & -b & 0 & -a & 0 & 0\\
0 &\frac{-a^2+b^2}{\sqrt{a^2+b^2}} & 0 & 0 & 0 & 0 & \frac{ab}{\sqrt{a^2+b^2}} & \frac{\sqrt{3}ab}{\sqrt{a^2+b^2}}\\
0 & -\frac{\sqrt{3} ab}{\sqrt{a^2+b^2}} & 0 & 0 & 0 & 0 & \frac{\sqrt{3} b^2}{2\sqrt{a^2+b^2}} & \frac{-2 a^2+b^2}{2\sqrt{a^2+b^2}}
\end{pmatrix}
\eeq
In the new basis the time-evolution can be solved from
\beq
\frac{d}{dt}\underline{{\lambda}}^{\prime T}=V'\underline{\lambda}^{\prime T},
\eeq
where $V'$ is a block-diagonal matrix given by
\beq
V'=BVB^T=
\begin{pmatrix}
V'_3 & & \\
& V'_4 &\\
&& V'_1
\end{pmatrix}
\eeq
and
\beq
V_3'=\begin{pmatrix}
0 & -\Delta & 0\\
\Delta & 0 & 2\epsilon\\
0 & -2\epsilon & 0
\end{pmatrix},\quad
V_4'=\begin{pmatrix}
0 & -\epsilon & 0 & 0\\
\epsilon & 0 & \Delta & 0\\
0 & -\Delta & 0 & -\epsilon\\
0 & 0 & \epsilon & 0
\end{pmatrix},\quad
V_1'=0
\eeq
with
\beq
\epsilon=\Omega_0\sqrt{a^2+b^2}.
\eeq
This result means that the time-evolution of the system can be analyzed
in terms of three separate vectors:
\beq
\underline{\Lambda}_3=(\lambda_1',\lambda_2',\lambda_3'),\quad
\underline{\Lambda}_4=(\lambda_4',\lambda_5',\lambda_6',\lambda_7'),\quad
\underline{\Lambda}_1=(\lambda_8'),
\eeq
Because of the antisymmetry of the $V'$-matrices, the lengths of these vectors are
conserved
\begin{align}
\label{Lambdas}
|\underline{\Lambda}_3|=\textrm{const.},\quad
|\underline{\Lambda}_4|=\textrm{const.},\quad
|\underline{\Lambda}_1|=\textrm{const.}
\end{align}
These are not independent quantities as they are tied together by the normalization
condition $|\underline{\Lambda}_3|^2+|\underline{\Lambda}_4|^2+
|\underline{\Lambda}_1|^2=|\underline{\lambda}'|^2=$const.
Therefore, in the presence of two monochromatic lasers and under two-photon resonance,
there are two conserved quantities in addition to the one arising from the conservation of $|\lambda'|$.
Due to unitary time-evolution $\underline{\lambda}'(t)=\Lambda_3(t)\oplus\Lambda_4(t)\oplus\Lambda_1(t)$
assumes only values which produce a positive density matrix $\rho(t)$.
Using (\ref{B}) the conserved quantities  (\ref{Lambdas}) can be expressed
in terms of the components of the original Bloch vector $\underline{\lambda}$.
For example, the time-independence of $|\Lambda_1|$ means that
\beq
\left(2\sqrt{3}ab\lambda_2(t)+\sqrt{3} b^2 \lambda_7(t)-(2a^2-b^2)\lambda_8(t)\right)^2=\textrm{const}\times (a^2+b^2)^2,
\eeq
where the value of the constant is determined by the initial values of $\lambda_2,\lambda_7$, and $\lambda_8$.
In \cite{Gottlieb82,Pegg85} it has been shown that similar approach allows two identify
two constants of motion in a three-level system under the assumption that
the Rabi frequencies are sinusoidally modulated, with a phase difference $\pi/2$ between them.
In this case the matrix determining the time-evolution is not block diagonal but enables nevertheless the time-evolution to be solved.

This approach has been extended to systems with more than three levels \cite{Hioe85}. If the Hamiltonian has a certain form,
obtained by defining a counterpart of the Hamiltonian (\ref{HHioe}) in the $n$ level case,
$n$ conserved quantities can be identified.

\subsection{The Bloch vector of two-qubit system}
We now study the Bloch vector description of a system consisting of two qubits, denoted by  $A$ and  $B$.
The Hilbert space $\mathcal{H}=\mathcal{H}_A\otimes\mathcal{H}_B=\mathbb{C}^2\otimes \mathbb{C}^2$ is
four dimensional and the density matrices can be expressed
using the the generalized $4\times 4$ Gell-Mann matrices (\ref{GGM}).
However, instead of using this basis, it is often advantageous to define the basis
in terms of the basis elements of the Lie algebra of $SU(2)\otimes SU(2)$.
This approach has been used in many articles, see, for example,
\cite{Fano83,Schlienz95,Aravind96,James01,Abouraddy02,Jaeger03,BK03,Theodorescu03,Altafini04,Jakob07}.
The latter choice for the basis allows to examine the
entanglement of two-qubit systems in a more natural way than the generalized Gell-Mann matrix basis.
We choose the basis $\{\hat{\lambda}_1,\hat{\lambda}_2,\ldots,\hat{\lambda}_{15}\}$ of the Lie algebra of $SU(2)\otimes SU(2)$ as
\begin{align}
\hat\lambda_i &=\frac{1}{\sqrt{2}}\sigma_i\otimes I_2,\quad i=1,2,3\\
\hat\lambda_i &=\frac{1}{\sqrt{2}}I_2\otimes\sigma_{i-3}\quad i=4,5,6\\
\hat\lambda_i &=\frac{1}{\sqrt{2}}\sigma_1\otimes\sigma_{i-6}\quad i=7,8,9\\
\hat\lambda_i &=\frac{1}{\sqrt{2}}\sigma_2\otimes\sigma_{i-9}\quad i=10,11,12\\
\hat\lambda_i &=\frac{1}{\sqrt{2}}\sigma_3\otimes\sigma_{i-12}\quad i=13,14,15.
\end{align}
This forms an orthogonal basis with respect to trace with the normalization given by
\beq
\Tr(\hat\lambda_i\hat\lambda_j)=2\,\delta_{ij}.
\eeq
The totally symmetric structure constants are
\begin{align}
\nonumber
&g_{147}=g_{158}=g_{169}=g_{24(10)}=g_{25(11)}=g_{26(12)}=g_{34(13)}=g_{35(14)}=g_{36(15)}\\
&=-g_{7(11)(15)}=g_{7(12)(14)}=g_{8(10)(15)}=-g_{8(12)(13)}=-g_{9(10)(14)}=g_{9(11)(13)}=\frac{1}{\sqrt{2}}.
\end{align}
These are needed in the calculation of the coefficients of the characteristic polynomial.
An arbitrary two-qubit state can be expressed as
\begin{align}
\rho &=\frac{1}{4}I_4 +\frac{1}{2}\sum_{i=1}^{15}\lambda_i\,\hat\lambda_i\\
\label{rho2qubit}
&=\frac{1}{2\sqrt{2}}\Big(\frac{1}{\sqrt{2}}I_4 +\sum_{i=1}^{3}\lambda_i\,\sigma_i\otimes I_2
+\sum_{i=1}^{3}\lambda_{i+3}\, I_2\otimes \sigma_i+\sum_{i=1}^{3}\sum_{j=1}^{3}\lambda_{j+3i+3}\,\sigma_i\otimes\sigma_j\Big),
\end{align}
where  $\underline{\lambda}$ is assumed to be such that $\rho$ is positive.
The reduced single-particle density matrices determined by this state are
\begin{align}
\rho_A=\Tr_B(\rho)=\frac{1}{2}I_2+\frac{1}{\sqrt{2}}\sum_{i=1}^3\lambda_i \sigma_i,
\quad \rho_B=\Tr_A(\rho)=\frac{1}{2}I_2+\frac{1}{\sqrt{2}}\sum_{i=1}^3\lambda_{i+3} \sigma_i
\end{align}
Here $\Tr_A$ denotes  trace over subsystem $A$, $\Tr_B$ is defined similarly.
The components of the Bloch vectors of the reduced states are uniquely obtained from the components of
the Bloch vector of $\rho$. The inverse does not hold; any information regarding the $\lambda_7,\lambda_8,\ldots,\lambda_{15}$ components
of the Bloch vector is missing from the reduces states $\rho_A$ and $\rho_B$.
The knowledge of the Bloch vectors of $\rho_A$ and $\rho_B$ does no allow to construct a {\it unique}
two-qubit Bloch vector which gives rise to $\rho_A$ and $\rho_B$.

As an example of the use of the parametrization we consider the Werner state for two qubits
\begin{equation}
\rho_W(x) = \frac{1-x}{4} I_4 + x S,
\end{equation}
where $x$ is a real parameter and $S$ is
\begin{eqnarray}
S = \frac{1}{2} \left(\begin{array}{cccc}
        0 & 0 & 0 & 0 \\
        0 & 1 & -1 & 0 \\
        0 & -1 & 1 & 0 \\
        0 & 0 & 0 & 0
            \end{array}\right).
\end{eqnarray}
We first determine the range of the parameter $x$ which ensures positivity of   $\rho_W(x)$, making sure that
$\rho_W(x)$ indeed is a density operator.  After this we study the separability of the Werner state $W(x)$.
We may write $W(x)$ as
\begin{eqnarray}
W(x) &=& \frac{1}{4} I_4 - \frac{x}{4}(\sigma_1 \otimes \sigma_1 +
        \sigma_2 \otimes \sigma_2 +\sigma_3 \otimes \sigma_3 )
            \nonumber  \\
       &=&\frac{1}{4}\left(\begin{array}{cccc}
    1-x &          0    &        0     &       0        \\
    0            & 1+x & -2x &       0        \\
    0            & -2x & 1+x &       0        \\
    0            &     0        &        0      &  1-x
           \end{array}\right).
\end{eqnarray}
The only non-zero components of the Bloch vector are $\lambda_7=\lambda_{11}=\lambda_{15}=-x/\sqrt{2}$.
A straightforward calculation gives the coefficients of the characteristic polynomial (\ref{a1234})
\begin{align}
1! a_1&=1\\
2! a_2&=\frac{3}{4}(1-x^2)\\
3! a_3 &=\frac{3}{8}(1-3x^2+2x^3)\\
4! a_4 &=\frac{3}{32}(1-6x^2+8x^3-3x^4)
\end{align}
These are all non-negative when $-1/3\leq x \leq 1$, which is therefore the range of $x$ corresponding
to physical states. Usually, however, the range of $x$ is taken to be $[0,1]$.
 We now examine the separability of $W(x)$. The density matrix $\rho$ of a composite
system $\mathcal{H}_A\otimes \mathcal{H}_B$ is called {\it separable} when it can be written
as a probabilistic mixture of tensor product states
\begin{align}
\rho=\sum_i p_i\rho_A^i\otimes\rho_B^i,\quad p_i\geq 0,\quad \sum_i p_i=1.
\end{align}
If a state is not separable it is {\it entangled}.
Detecting separability is in general a very complicated problem, but in some special
cases it can be done easily.
A simple way to test the separability of a two-qubit system is to use the Peres-Horodecki,
or positive partial transposition, criterion \cite{Peres96,Horodecki96}. It provides
a necessary and sufficient condition for separability for $2\times 2$ and $2\times 3$ dimensional systems.
According to this criterion, a state $\rho$ is separable if the operator obtained by transposing
the density operator of the subsystem $A$ or $B$ is a positive operator. When the state of
the second subsystem is transposed, (\ref{rho2qubit}) becomes
\begin{align}
\rho^{\textrm{pt}}
&=\frac{1}{2\sqrt{2}}\Big(\frac{1}{\sqrt{2}}I_4 +\sum_{i=1}^{3}\lambda_i\,\sigma_i\otimes I_2
+\sum_{i=1}^{3}\lambda_{i+3}\, I_2\otimes \sigma_i^T+\sum_{i=1}^{3}\sum_{j=1}^{3}\lambda_{j+3i+3}\,\sigma_i\otimes\sigma_j^T\Big)\\
&=\frac{1}{2\sqrt{2}}\Big(\frac{1}{\sqrt{2}}I_4 +\sum_{i=1}^{3}\lambda_i^{\textrm{pt}}\,\sigma_i\otimes I_2
+\sum_{i=1}^{3}\lambda_{i+3}^{\textrm{pt}}\, I_2\otimes \sigma_i+\sum_{i=1}^{3}\sum_{j=1}^{3}\lambda_{j+3i+3}^{\textrm{pt}}\,\sigma_i\otimes\sigma_j\Big),
\end{align}
where $\lambda_5^{\textrm{pt}}=-\lambda_5,\lambda_8^{\textrm{pt}}=-\lambda_8,
\lambda_{11}^{\textrm{pt}}=-\lambda_{11},\lambda_{14}^{\textrm{pt}}=-\lambda_{14}$ and $\lambda_i^{\textrm{pt}}=\lambda_i$ for
the rest of the components. When partial transpose is applied to the Werner state we get
\begin{eqnarray}
W_{\textrm{pt}}(x) &=& \frac{1}{4} I_4 - \frac{x}{4}(\sigma_1 \otimes \sigma_1 -
        \sigma_2 \otimes \sigma_2 +\sigma_3 \otimes \sigma_3 )
            \nonumber  \\
\label{Wpt}
       &=& \frac{1}{4}\left(\begin{array}{cccc}
    1-x&          0    &        0     &   -2x      \\
    0            & 1+x &       0      &       0        \\
    0            &       0       &    1+x &       0        \\
-2x     &      0        &        0      &  1-x
           \end{array}\right).
\end{eqnarray}
The calculation of the coefficients $a_i^{\textrm{pt}}$ of the characteristic polynomial of $W_{\textrm{pt}}(x)$ shows
that $a_1^{\textrm{pt}}$ and $a_2^{\textrm{pt}}$ are the same than for $W(x)$ and
\begin{align}
3! a_3^{\textrm{pt}} &=\frac{3}{8}(1-3x^2-2x^3)\\
4! a_4^{\textrm{pt}} &=\frac{3}{32}(1-6x^2-8x^3-3x^4).
\end{align}
These differ from $a_3$ and $a_4$ by the sign of the $x^3$ -term.
These coefficients are non-negative when $-1\leq x\leq 1/3$. Therefore the Werner
state $W(x)$ is separable when $-1/3\leq x \leq 1/3$ and entangled when $1/3 < x\leq 1$.
Because $W(x)$ is obtained by partially transposing $W_{\textrm{pt}}(x)$, the state $W_{\textrm{pt}}(x)$ is separable
when $-1/3\leq x\leq 1/3$ and entangled when $-1\leq x<-1/3$.

\subsection{The polarization operator basis}
Our definition of the Bloch vector with respect to the generalized Gell-Mann
matrices is not the only possible. Although changing the basis will change the Bloch vector representation of the states,
it does not affect the structure of the set of density matrices in general.
One possible way to choose the basis is to use the polarization operators, also known as
spherical tensor operators.
The concept of a polarization operator appears in the quantum mechanical theory of angular momentum
and is thoroughly discussed in the literature,see, for example, \cite{Biedenharn,Varshalovich}.
When the angular symmetries of the
system are important it is convenient to expand $\rho$ using the polarization operators.
Many examples of such  systems can be found in \cite{Blum96}.
We will describe the polarization operators only briefly.
In the following discussion we define the basis and
characterize some of the properties of the set of Bloch vectors corresponding to physical states.
This subsection is based on \cite{Kryszewski06,Bertlmann08}.

The polarization operators pertaining to an $n$-level system are defined as
\begin{equation} \label{defpo}
    T_{LM} \;=\; \sqrt{\frac{2L+1}{2s+1}} \sum_{k,l =1}^n C^{s m_k}_{s
    m_l , \,LM} \,|k \rangle \langle l | \,,
\end{equation}
where the indices have the properties
\begin{align}
    & s = \frac{n-1}{2} \,, & \nonumber\\
    & L = 0,1, \ \ldots \ ,2s \,, & \nonumber\\
    & M = -L, -L+1, \ldots, L-1, L \,, & \nonumber\\
    & m_1 = s, \ m_2 = s-1, \ldots ,m_n = -s \,.
\end{align}
The coefficients $C^{s m_k}_{s m_l , \,LM}$ are identified with the usual Clebsch--Gordan
coefficients $C^{j m}_{j_1 m_1 , \,j_2 m_2}$ of the theory angular momentum.
For $L=M=0$ the polarization operator is proportional to the identity matrix
\begin{equation} \label{po00}
T_{00} = \frac{1}{\sqrt{n}} I_n.
\end{equation}
All other polarization operators are traceless, but they are {\it not} Hermitian in general.
Due to the symmetry properties of the Clebsch--Gordan
coefficients they satisfy the orthogonality relation
\begin{equation}\label{orthogonal-pob}
\Tr( T_{L_1 M_1}^\dag T_{L_2 M_2}) = \delta_{L_1 L_2}\delta_{M_1 M_2}
\end{equation}
and they have the property
\begin{equation}
T_{LM}^\dag=(-1)^M T_{L-M}.
\end{equation}
The former equation ensures that the polarization operators form an orthonormal basis.
Any density matrix can be written using this basis as
\begin{equation} \label{bvpob-d}
    \rho =\ \frac{1}{n} I_n + \sum_{L=1}^{2s} \sum_{M=-L}^L \lambda_{LM}^{\textrm{po}} T_{LM}
     = \frac{1}{n} I_n + \underline{\lambda}^{\textrm{po}} \cdot\underline{T}
\end{equation}
with the Bloch vector in the polarization operator basis given by
$\underline{\lambda}^{\textrm{po}} =(\lambda_{1-1}^{\textrm{po}},\lambda_{10}^{\textrm{po}},\lambda_{11}^{\textrm{po}},
\lambda_{2-2}^{\textrm{po}},\lambda_{2-1}^{\textrm{po}},\lambda_{20}^{\textrm{po}},...,\lambda_{(n-1)(n-1)}^{\textrm{po}})$, where the components are
ordered and given by $\lambda_{LM}^{\textrm{po}} = \Tr(T_{LM}^\dagger \rho)$. In general, the
components $\lambda_{LM}^{\textrm{po}}$ are complex since the polarization operators are not
Hermitian. The hermiticity of the density matrix, $\rho=\rho^*$, forces the components
of the Bloch vector to fulfill the condition
\begin{equation}
\label{LMequation}
\lambda_{LM}^{\textrm{po}}=(-1)^M \bar{\lambda}_{L-M}^{\textrm{po}}
\end{equation}
In particular, the components $\lambda_{L0}^{\textrm{po}}$ are real.
In order for $\rho$ to describe a physical state it also has to be positive.
As in the case of the Gell-Mann matrix basis, the positivity of $\rho$ can be checked
using (\ref{char-poly-n}) and (\ref{ev-pos}).
The coefficients $a_j$ of the characteristic polynomial (\ref{char-poly-n})
are related to the trace invariants through the equation
\begin{equation}
j a_j=\sum_{m=1}^j (-1)^{m-1} a_{j-m}\Tr(\rho^m).
\end{equation}
General expression for the trace invariants can be calculated to be
\begin{eqnarray}
\nonumber
\Tr(\rho^k)&=&\frac{1}{n^{k-1}}+\frac{k(k-1)}{2n^{k-2}}|\underline{\lambda}^{\textrm{po}}|^2
+\sum_{m=3}^k {k \choose m} \frac{\Tr\left[(\underline{\lambda}^{\textrm{po}} \cdot\underline{T})^m\right]}{n^{k-m}},\\
\Tr\left[(\underline{\lambda}^{\textrm{po}} \cdot\underline{T})^m\right]
&=& \displaystyle\sum_{L_1=1}^{n-1}\sum_{M_1=-L_1}^{L_1}\cdots \sum_{L_m=1}^{n-1}\sum_{M_m=-L_m}^{L_m}
\lambda^{\textrm{po}}_{L_1 M_1}\cdots \lambda^{\textrm{po}}_{L_m M_m}
\Tr(T_{L_1 M_1}\cdots T_{L_m M_m}).
\end{eqnarray}
The traces can be calculated using the equations presented in \cite{Varshalovich}.
Combining the results we find that the coefficients of the characteristic polynomial are
\begin{equation}
j a_j =a_{j-1}+\sum_{k=2}^j (-1)^{k-1} a_{j-k}\left[\frac{1}{n^{k-1}}+\sum_{m=2}^k{k\choose m}
\frac{\Tr\left[(\underline{\lambda}^{\textrm{po}} \cdot\underline{T})^m\right]}{n^{k-m}}\right].
\end{equation}
The first three coefficients are $a_0=a_1=1$ and
\begin{equation}
2 a_2= 1-\frac{1}{n}-|\underline{\lambda}^{\textrm{po}}|^2.
\end{equation}
Therefore a necessary condition for $\rho$ to be a positive operator is
that Bloch vectors lie within a hypersphere of radius $|\underline{\lambda}^{\textrm{po}}| \leq\sqrt{(n-1)/n}$.
In the case of a two-level system ($n=2$) the density matrix becomes
\begin{equation}
\rho =\frac{1}{2} I_2+(\alpha +i\beta)T_{11}-(\alpha-i\beta) T_{1-1}+\gamma T_{00},
\end{equation}
where $\alpha, \beta,\gamma$ are real and we have defined
$\lambda_{11}^{\textrm{po}}=\alpha +i\beta$, $\lambda_{00}^{\textrm{po}}=\gamma$ and
used (\ref{LMequation}). The set of physical states is now given by
$Q_2=\{(\alpha,\beta,\gamma)\in \mathbb{R}^3\, |\, 2(\alpha^2+\beta^2)+\gamma^2\leq 1/2\}$.
The surface of this set corresponds to pure states and is a prolate spheroid.

As in the case of the Bloch vector given in the generalized Gell-Mann basis,
for $n\geq 3$ the structure of the set of physical states becomes very complicated \cite{Kryszewski06}.
Nevertheless, pure states are on the surface, mixed ones lie within the volume and the maximally mixed state corresponds to
$|\underline{\lambda}^{\textrm{po}}|=0$.

\section{The coset parametrization}\label{sec:Coset}
This section presents a short summary of the article \cite{Ak07}. If $D=D_n(\lambda_1,\ldots,\lambda_n)$ is the diagonal matrix of eigenvalues in the spectral representation (\ref{spec-rep}) of a density matrix $\rho$, we denote by $D\pr$
the commutant of $D$ in $U(n)$, i.e., the set of all matrices $U\in U(n)$ such that
$UD=DU$. Now, if (\ref{spec-rep}) holds for some $U \in U(n)$ and $V$ is some other unitary matrix such that $V U^{-1} \in D\pr$ then one has
\beq \label{specprep-nonunique}
\rho =U^* D U = V^* D V, \qquad V U^{-1} \in D\pr,
\eeq
and conversely.

In the case of nondegenerate spectrum the commutant $D\pr$ is easily determined:
\beq \label{commutant-nondeg}
D\pr =T^n = U(1)\otimes \cdots \otimes U(1), \quad n \;\rm{factors},
\eeq
and thus one can say
\beq \label{coset1}
\rho = \Omega ^* D \Omega, \quad \Omega \in U(n)/{T^n}.
\eeq
According to the book \cite{Gi74} elements $U \in U(n)$ can be factored in the following way:
\beq \label{unitary-fact}
U =\Omega_n \Omega_{n-1} \cdots \Omega_2 \Omega_1
\eeq
with $\Omega_1 \in T^n$ and
\beq \label{coset-elements}
\Omega_k \in \frac{U(k)\otimes T^{n-k}}{U(k-1)\otimes T^{n-k+1}},\quad k=2,\ldots,n \; .
\eeq
Typical coset representatives $\Omega_k$ are of the form
\beq \label{coset-rep}
\renewcommand{\arraystretch}{1.2}
\left( \begin{array}{c|c}
SU(k)/U(k-1) &\0\\ \hline
\vspace*{3mm}
\0^T& I_{n-k}\\
\end{array}
\right)
\eeq
where $\0$ is the $k \times (n-k)$ zero matrix and $\0^T$ its transpose while $I_{n-k}$ denotes the $n-k$ dimensional unit matrix.

After a brief sketch of the general case the article \cite{Ak07} proceeds to discuss in some detail the cases $n=2$ and $n=3$ (and the Bures metric and the $n=2$ state space). A more systematic approach and an explicit realization of the factorization (\ref{unitary-fact}) is given in the following section. Therefore no further details are presented here.

\section{The Jarlskog parametrization}\label{sec:Jarlskog}
Like the coset parametrization also the Jarlskog parametrization starts from the spectral representation (\ref{spec-rep}) of a density matrix. Accordingly, we begin with a concrete version of the factorization of a unitary matrix $U_n \in SU(n)$ in terms of certain basic unitary matrices as in (\ref{unitary-fact}) following \cite{Ja06}. Since this factorization is based on the use of canonical coordinates of second kind in the Lie group $SU(n)$, a more accurate name for the parametrization would be the ``canonical coordinate parametrization". Some preliminary investigations into this problem have been given in the article \cite{BP08}.
After having defined the Jarlskog parametrization of unitary matrices, we show  how it can be used to parametrize density matrices. As an application of this parametrization we construct a general density matrix of a two-level system. We also show that the Jarlskog parametrization can be straightforwardly extended to composite systems and illustrate
this by constructing some two-qubit states.

\subsection{Jarlskog's recursive parametrization of unitary matrices}
Recall that the Lie group $U(n)$ is connected but not
simply connected and that every $U_n \in U(n)$ has a
determinant of absolute value $1$. It follows that every
$U_n \in U(n)$ has a determinant $\det U_n =e^{i\alpha} $
and thus can be written as
$$U_n = D_n(e^{i\alpha},1,\ldots,1)U_n\pr,\quad \alpha \in \R,
\quad U_n\pr \in SU(n)$$ where $SU(n)$ denotes the Lie
group of unitary matrices of determinant $1$. Obviously,
the diagonal matrices $D_n(e^{i\alpha},1,\ldots,1)$ and
$D_n(\lambda_1,\ldots, \lambda_n)$   commute and hence in
(\ref{spec-rep}) we can restrict ourselves to unitary
matrices in $SU(n)$, i.e., $\rho_n \in \cald_n$ if, and only
if, \beq \label{eq:reduction2} \rho_n = U_n^*
D_n(\lambda_1,\ldots, \lambda_n) U_n, \;
 U_n \in SU(n). \eeq
where the  eigenvalues $\lambda_j$ satisfy (\ref{ev-s}) and (\ref{ev-dens}).
This reduces our problem to that of finding a
parametrization of unitary matrices of determinant $1$.

In order to provide the necessary background for this parametrization of unitary
matrices and the important recursion formula we follow Fujii \cite{Fu05} to explain
the origin of the basic building blocks of this parametrization. He  observed
that the Jarlskog parametrization of unitary matrices is obtained by using
canonical coordinates of the second kind for the Lie group $SU(n)$
(see, for instance, \cite{Va84}).

Recall that $SU(n)$ is the (simply) connected component of
the unit element of $U(n)$ and thus is the image of its Lie
algebra $su(n)$ under the exponential map. The Lie algebra
 $su(n)$ consists of all skew-adjoint $n\times n$ matrices
 $X$. Such matrices have the form
 \beq \label{skewadj-mat}
 X = \begin{pmatrix}
 i \alpha_1 & z_{12} & z_{13} & \cdots & z_{1,n-1} &z_{1n} \\
 -\bar{z}_{12} & i \alpha_2 & z_{23} & \cdots &
 z_{2,n-1} & z_{2n} \\
 -\bar{z}_{13} & -\bar{z}_{23} & i \alpha_3 &
 \cdots & z_{3,n-1} & z_{2n} \\
 \vdots & \vdots & \vdots & \ddots & \vdots & \vdots \\
 -\bar{z}_{1,n-1} & -\bar{z}_{2,n-1} &
 \bar{z}_{3,n-1} & \cdots & i \alpha_{n-1} & z_{n-1,n}  \\
 -\bar{z}_{1,n} & -\bar{z}_{2,n} & -\bar{z}_{3,n}
 & \cdots & -\bar{z}_{n-1,n} & i \alpha_n
 \end{pmatrix} \eeq
where the $\alpha_j$'s are real numbers and the
$z_{jk}$ are complex numbers which can be chosen
independently.
Such a matrix has the natural decomposition
\beq \label{dec-skewmat} X = X_1 + X_2 + \cdots + X_j +
\cdots + X_n \eeq where $X_1$ is the diagonal matrix with
the diagonal entries $i\alpha_1, \ldots, i\alpha_n$: \beq
\label{diag-mat0} X_1 =D_n(i \alpha_1, \ldots, i \alpha_n)
\eeq and where for $ j=2,\ldots,n$ the matrix $X_j$ is the
matrix which has in column $j$ the column vector \beq
\label{c-vector} |z_j\ran =
\begin{pmatrix}
z_{1j} \\ z_{2j} \\ \vdots \\ z_{j-1,j} \\
\end{pmatrix}  \in \C^{j-1}     \eeq
and in row $j $ the row vector \beq \label{r-vector} -
\lan z_j| = (- \bar{z}_{1j},-\bar{z}_{2j}, \ldots,
-\bar{z}_{j-1,j} ) \eeq as entries. All other entries of $X_j$ are zero.

The canonical coordinates of the second kind for $SU(n)$ are given by
\beq \label{canonical coordinates}
su(n) \ni X=X_1 + X_2 + \cdots + X_j + \cdots + X_n \longrightarrow e^{X_1} e^{X_2} \cdots e^{X_j} \cdots
e^{X_n} \in SU(n).
\eeq

The exponential $e^{X_1}$  is easily  calculated:
\beq \label{eq:An1} A_{n,1}=e^{X_1}=
A_{n,1}(\alpha_1,\ldots,\alpha_n)=D_n(e^{i\alpha_1},
\ldots, e^{i\alpha_n}), \quad \alpha_j \in \R. \eeq
For $j=2,\ldots,n-1$ we write
\beq \label{Xj}
\renewcommand{\arraystretch}{1.2}
X_j = \left( \begin{array}{cc}
K_j & \0 \\
\0 &\; \0_{n-j}
\end{array} \right),
\quad K_j= \left( \begin{array}{cc}
\0_{j-1}\; & |z_j\rangle \\
-\langle z_j| & \0
\end{array} \right),
\eeq
where $\0_k$ denotes the $k \times k$ zero matrix and $\0$ indicates that the remaining entries of the matrix are zero.
Naturally $X_n=K_n$. Thus we get
\beq \label{exp1}
e^{X_j} = \begin{pmatrix}
e^{K_j} & \0 \\ \0 & I_{n-j}
\end{pmatrix}, \quad j =2,\ldots, n-1, \quad e^{X_n}= e^{K_n}
\eeq
where $I_k$ denotes the $ k \times k$ unit matrix.
Using
$$
K_j^2 = -\begin{pmatrix} |z_j\ran \lan z_j| & 0\\ \0 &\;\; \lan z_j|z_j \ran \end{pmatrix}, \quad K_j^3 = -\lan z_j|z_j\ran K_j
$$
one can calculate the exponentials and finds for $j=2, \ldots,n$, $|z_j|= \sqrt{\lan z_j|z_j\ran}$,
\beq \label{exp2}
e^{K_j} = I_j +(1-\cos{|z_j|}) \frac{1}{|z_j|^{2}}K_j^2 + \sin{|z_j|} \frac{1}{|z_j|} K_j \; .
\eeq
If one introduces the unit vector $|\tilde{z}_j\ran= \frac{1}{|z_j|} |z_j\ran \in \C^{j-1}$ one can rewrite this exponential as
\beq \label{exp2}
e^{K_j} = I_j -(1-\cos{|z_j|}) \begin{pmatrix}
|\tilde{z}_j\ran \lan\tilde{z}_j| & \;0 \\ \0 & \;1 \end{pmatrix} +
\sin{|z_j|} \begin{pmatrix}
\0_{j-1} & \; |\tilde{z}_j\ran \\ -\lan\tilde{z}_j| & 0
\end{pmatrix}.
\eeq
The right hand side of (\ref{exp2}) is a unitary matrix $V_{n,j}$ which  is often written as
\beq
\label{eq:Vnj} V_{n,j} = \begin{pmatrix}
I_{j-1} - (1-c_j)|\tilde{z}_j\ran\lan \tilde{z}_j| & \;\; s_j |\tilde{z}_j\ran \\
-s_j \lan\tilde{z}_j| & c_j
\end{pmatrix},
\quad c_j=\cos{|z_j|},\quad s_j =\sin{|z_j|} \; .
\eeq
Since the adjoint (and the inverse) of the matrix $e^{K_j}$ is $e^{-K_j}$, the adjoint and the inverse of the matrix $V_{n,j}$ is given by (\ref{eq:Vnj}) with $\tilde{z}_j$
replaced by $-\tilde{z}_j$.

 The Jarlskog matrices $A_{n,j}=e^{X_j}$, $j=1,\ldots,n$ (see \cite{Ja05,Ja06}) thus are of the form
\beq \label{eq:Anj} A_{n,j} =\begin{pmatrix}
V_{n,j} & \0 \\ \0 & I_{n-j} \end{pmatrix},
\quad j=2,\ldots,n-1, \quad A_{n,n} =V_{n,n}.
\eeq

A generic element $U_n \in SU(n)$  therefore has  the factorization
 \beq \label{exp-explicit} U_n
=e^{X_1} e^{X_2}\cdots e^{X_n}= A_{n,1} A_{n,2} \cdots
A_{n,n} \eeq
or equivalently
\beq \label{exp-explicit2} U_n
=e^{X_n} e^{X_{n-1}}\cdots e^{X_1}= A_{n,n} A_{n,n-1} \cdots
A_{n,1}. \eeq
Observe that by construction the Jarlskog matrix $A_{n,j}$ is defined in terms of the following set of parameters
$$\theta_j \geq 0, \quad z_j \in S(\C^{j-1})$$
where we changed notation: $\theta_j$ for $|z_j|$ and $z_j$ for $\tilde{z}_j$,
and where $S(\C^{j-1})$ denotes the unit sphere in $\C^{j-1}$. We indicate this by writing $A_{n,j}=A_{n,j}(\theta_j,z_j)$. The special structure of the matrix $A_{n,j}$ and the properties of the trigonometric functions imply that on the parameter set
\beq \label{pnj}
P_{n,j}= \{(\theta_j,z_j):\; 0 \leq \theta_j \leq \pi/2,\, z_j \in S(\C^{j-1})\}
\eeq
the mapping $(\theta_j,z_j) \longrightarrow A_{n,j}$ is injective, i.e., if $(\theta_j,z_j), (\theta_j\pr,z_j\pr) \in P_{n,j}$ and $A_{n,j}(\theta_j,z_j)=
A_{n,j}(\theta_j\pr,z_j\pr)$ then $\theta_j =\theta_j\pr$ and $z_j = z_j\pr$.

According to (\ref{skewadj-mat}) or (\ref{exp-explicit}), (\ref{exp-explicit2}) and (\ref{pnj}) this factorization describes a generic element $U_n \in SU(n)$ in terms of $n^2$ real parameters.

Notice that the matrices $V_{n,j}$ do not depend explicitly on the dimension $n$ and that we can write, for $j=2,\ldots,n-1$,
\beq \label{J-matrix}
A_{n,j} = \begin{pmatrix}
\tilde{V}_{n,j} & \0 \\ \0 & 1 \end{pmatrix}, \quad  \tilde{V}_{n,j} = \begin{pmatrix}
V_{n,j} & \0 \\ \0 &\;\; I_{n-j-1} \end{pmatrix} . \eeq
It follows that
\beq \label{product-lo}
A_{n,1} \cdots A_{n,n-1} = \begin{pmatrix}
\tilde{V}_{n,1} \cdots \tilde{V}_{n,n-1} & \0 \\ \0 & 1 \end{pmatrix} . \eeq
The analysis presented above shows that $\tilde{V}_{n,1} \cdots \tilde{V}_{n,n-1}$,
up to a factor in $T^{n-1}$, gives the generic factorization of $(n-1)\times (n-1)$
special unitary matrices and hence, from (\ref{exp-explicit2}),
we get the important recursion relation
\beq \label{rec-rel}
U_n =\begin{pmatrix} U_{n-1} & \0 \\ \0 & 1 \end{pmatrix} A_{n,n}=\begin{pmatrix} U_{n-1} & \0 \\ \0 & 1 \end{pmatrix}
\begin{pmatrix}
I_{n-1} - (1-c_n)|{z}_n\ran\lan {z}_n| & \;\; s_n |{z}_n\ran \\
-s_n \lan {z}_n| & c_n
\end{pmatrix}
\eeq

Next we address the question of injectivity of the Jarlskog parametrization of $U_n \in SU(n)$ modulo elements in $T^n$, i.e., the parametrization (\ref{exp-explicit}) without the factor $A_{n,1} \in T^n$.  Then this parametrization uses the parameter set
\beq \label{Un-param-set}
P_n =  \{(\theta_j,z_j):\, (\theta_j,z_j) \in P_{n,j},\, j=2,\ldots,n\}
\eeq
where we have taken  (\ref{exp-explicit})  and (\ref{pnj}) into account. In order to prove injectivity of the map
\beq \label{Un-injectivity}
(\theta_2,z_2;\theta_3,z_3; \ldots;\theta_n,z_n) \longrightarrow U_n =A_{n,2}(\theta_2,z_2)A_{n,3}(\theta_3,z_3) \cdots A_{n,n}(\theta_n,z_n),
\eeq
where $(\theta_2,z_2;\theta_3,z_3; \ldots;\theta_n,z_n) \in P_n$,
we proceed by induction with respect to the order $n$. Since $(\theta_j,z_j) \longrightarrow A_{n,j}$ is injective on $P_{n,j}$, the map (\ref{Un-injectivity})
is injective for $n=2$. Now assume that for some $n >2$ this map is injective
for the orders $k \leq n-1$. We now show that the map (\ref{Un-injectivity}) then is injective for the order $k=n$.

Suppose that for $U_n,U_n\pr \in SU(n)$ we have $U_n\pr = U_n$ where $U_n\pr =
U_n(\theta_2\pr,z_2\pr;\theta_3\pr,z_3\pr,; \ldots;\theta_n\pr,z_n\pr)$. In the recursion formula
(\ref{rec-rel}) for $U_n$ we abbreviate $I_{n-1} -(1-c_n)|z_n\rangle\langle z_n|$ with $B_{n-1}$
and similarly for $U_n\pr$ and $B_{n-1}\pr$. If we calculate the matrix product in the recursion relation,
the identity $U_n\pr = U_n$ reads
\beq \label{rec-ident}
\begin{pmatrix}
U_{n-1}\pr B_{n-1}\pr \quad & s_n\pr U_{n-1}\pr |z_n\pr\ran \\
-s_n\pr \lan z_n\pr| & c_n\pr  \end{pmatrix} = \begin{pmatrix} U_{n-1} B_{n-1}\quad
& s_n U_{n-1} |z_n\ran \\ -s_n \lan z_n| & c_n  \end{pmatrix}
\eeq
where we used the abbreviations $s_n\pr =\sin{\theta_n\pr}$ and $c_n\pr =\cos{\theta_n\pr}$.
Thus $c_n\pr =c_n$; since $\theta_n,\theta_n\pr \in [0,\pi/2]$ we conclude $\theta_n\pr =\theta_n$
and therefore $s_n\pr =s_n$. The identity $-s_n\pr \lan z_n\pr| = -s_n \lan z_n|$ now implies
$\lan z_n\pr|= \lan z_n|$ and $|z_n\pr\ran=|z_n\ran$. Next we use the identity
$s_n \pr U_{n-1}\pr |z_n\pr\ran = s_n U_{n-1}|z_n\ran$ to conclude
 $U_{n-1}\pr |z_n\pr\ran = U_{n-1}\pr |z_n\ran = U_{n-1}|z_n\ran$. Finally we use the identity
 $U_{n-1}\pr B_{n-1}\pr =U_{n-1} B_{n-1}$ to get
 $$U_{n-1}\pr -(1-c_n\pr)U_{n-1}\pr|z_n\pr\ran \lan z_n\pr|=U_{n-1} -(1-c_n)U_{n-1}|z_n\ran\lan z_n|$$
and from the identities established above we find $U_{n-1}\pr = U_{n-1}$.
The induction hypothesis implies $(\theta_j \pr, z_j\pr) = (\theta_j,z_j)$ for $j=2,\ldots, n-1$, and we conclude.

\subsection{Jarlskog parametrization}
Since diagonal matrices commute, (\ref{eq:reduction2})
implies  the following parametrization of a generic density
matrix $\rho_n \in \cald_n$: \beq \label{eq:dens-matrix-param}
\rho_n =U_n^* D_n(\lambda_1,\ldots,\lambda_n) U_n =  A_{n,n}^*\cdots
A_{n,2}^*D_n(\lambda_1,\ldots,\lambda_n)A_{n,2} \cdots A_{n,n} \eeq where the matrices
$A_{n,j}$ and the matrix $D_n$ are parametrized as described above. Recall that for each
$j=2,\ldots,n$ both the parameter set $P_{n,j}$ for the matrix $A_{n,j}$
and the concrete form of this matrix have been given explicitly.
The number of parameters is easily calculated: There are $n^2 - n$ real parameters
for the product $A_{n,2} \cdots A_{n,n}$ and $n -1$ parameters representing
the eigenvalues $\lambda_1, \ldots, \lambda_n$ subject to the normalization condition
$\sum_{j=1}^n \lambda_j =1$. This gives $n^2 - 1$ independent real parameters as in the
Bloch vector parametrization.

If we introduce the set
\beq \label{ev-param}
\Lambda_n = \{\und{\lambda}=(\lambda_1, \ldots, \lambda_n)\in \R^n : \; \lambda_1 \geq \lambda_2 \geq \cdots \geq \lambda_n \geq 0, \;\sum_{j=1}^n \lambda_j =1 \}
\eeq
we can specify the parameter set $Q_n$ of this parametrization as
\beq \label{parameter-set-Jarlskog}
Q_n = \{ (\und{\lambda}, \theta_j, z_j):\, \und{\lambda} \in \Lambda_n, \, \theta_j \in [0,\pi/2], z_j \in S(\C^{j-1}), j=2,\ldots, n \}.
\eeq

Our recursion relation (\ref{rec-rel}) for unitary matrices leads to an interesting
recursion relation for density matrices in different dimensions. Denote the product
$A_{n,2} \cdots A_{n,n-1}$ by $U_{n-1}$. Then, using (\ref{rec-rel}), we can deduce
from (\ref{eq:dens-matrix-param}) the recursion formula
\begin{align} \label{rec-rel-dens}
\rho_n &= A_{n,n}^* \begin{pmatrix} U_{n-1}^* & \0 \\ \0 &1 \end{pmatrix}
\begin{pmatrix} D_{n-1}(\lambda_1, \ldots, \lambda_{n-1} ) & \0 \\ \0 & \lambda_n
\end{pmatrix}
\begin{pmatrix} U_{n-1} & \0 \\ \0 & 1 \end{pmatrix} A_{n,n}\\
&=A_{n,n}^* \begin{pmatrix} \rho_{n-1} & \0 \\ \0 & \lambda_n \end{pmatrix} A_{n,n}
\end{align}
where
\beq \label{red-dens-matrix}
\rho_{n-1} =U_{n-1}^* D_{n-1}(\lambda_1, \ldots, \lambda_{n-1} )U_{n-1}
\eeq
is a positive matrix in dimension $n-1$ with trace $\Tr{\rho_{n-1}} = 1- \lambda_n$.

While in the case of the Bloch vector parametrization the question of uniqueness was
quite easy to answer, it is fairly complicated in the case of the Jarlskog parametrization
and only partial answers are known. Given $\und{\lambda} \in \Lambda_n$, consider
the diagonal matrix $D_n(\und{\lambda})$ of eigenvalues and the commutant of this matrix in $SU(n)$,
\beq \label{commutant-diag}
D_n(\und{\lambda})\pr = \{ V \in SU(n):\, V D_n(\und{\lambda}) = D_n(\und{\lambda}) V \}.
\eeq
Clearly this commutant depends on $\und{\lambda} \in \Lambda_n$. As mentioned in the previous section,
in the case of a non-degenerate spectrum, i.e., if $\lambda_1 > \lambda_2 > \cdots > \lambda_n \geq 0$,
then this commutant is easily determined
\beq \label{commutant-nondegenerate}
D_n(\und{\lambda})\pr =\{V=D_n(e^{\ii \alpha_1}, \ldots, e^{\ii \alpha_n}):\, \sum_{j=1}^n \alpha_j =0\}.
\eeq

Next consider the case that for some $1\leq k <n$ one has $\lambda_1 > \lambda_2 > \cdots \lambda_k >0$
while $\lambda_j =0$ for $j=k+1, \ldots, n$. This case can be reduced to the previous one by
restricting the density matrix to the subspace spanned by the first $k$ eigenvectors.

In general one can not exclude degeneracy in the spectrum of a density matrix $\rho_n$;
then one has to consider eigenvalues $\lambda_j$ with multiplicity
$m_j$, $j=1,2,\ldots, k$, $k <n$, $\lambda_1 > \lambda_2 > \cdots \lambda_k$, $m_1 + \cdots + m_k =n$.
For this kind of a spectrum the commutant (\ref{commutant-diag}) is
\beq \label{com-deg-spec}
D_n(\und{\lambda})\pr = \{V = \begin{pmatrix}
V_1 & \0 & \cdots & \0\\ \0 & V_2 & \cdots & \0 \\
\vdots & \vdots  & \cdots & \vdots \\
\0 & \0 & \cdots &V_k
\end{pmatrix} : \; V_j \in U(m_j),\; \det {V}=\prod_{j=1}^k\det{V_j}=1 \}.
\eeq
In order to establish an injective parametrization of density matrices in this case one
has to determine the Jarlskog parametrization of elements $U_n \in SU(n)$ modulo
elements in $D_n(\und{\lambda})\pr$. This is not yet known.

\subsection{Simple examples}
As has been explained in section 2, the structure of the set of Bloch vectors is complicated when $n>2$.
In order to overcome this other parametrizations can be used.
To get some inspiration on how to proceed we have a new look at the Bloch-vector
parametrization for $n=2$. In the parametrization (\ref{d2-map2})
we again take $\theta =2 \vartheta$ and write the relations (\ref{roots-2d}) for
the eigenvalues as $|\und{\lambda}|= x_1 -x_2$
and $1=x_1 + x_2$. Then, with the abbreviations $c= \cos\vartheta$ and $s=\sin\vartheta$,
the parametrization (\ref{d2-map2}) can be written as
\beq \label{2-par}
\begin{pmatrix} x_1 c^2 + x_2 s^2 & (x_1 -x_2) cs e^{- \ii \phi} \\
(x_1 -x_2) cs e^{ \ii \phi} &  x_2 c^2 + x_1 s^2 \end{pmatrix}, \quad (x_1,x_2,\vartheta,\phi) \in
\hat{Q}_2
\eeq
where the parameter set $\hat{Q}_2$ has the form
\beq \label{p-set-2} \hat{Q}_2= \{(x_1,x_2,\vartheta,\phi)\in \R^4:\,
0\leq x_1, 0 \leq x_2 ,\;x_1 +x_2 =1, \; 0\leq \vartheta \leq \pi/2,\, 0\leq \phi < 2 \pi\}.
\eeq
In this form we can recognize the parametrization (\ref{2-par}) to be the following product of 5 matrices:
\beq \label{2-par-factors}
\begin{pmatrix} e^{-\ii \phi/2} & 0 \\ 0 & e^{\ii \phi/2} \end{pmatrix}
 \begin{pmatrix} c & \;-s\\ s& c \end{pmatrix}  \begin{pmatrix} x_1 & 0 \\ 0 & x_2  \end{pmatrix}
 \begin{pmatrix} c & s\\ -s\;& c \end{pmatrix} \begin{pmatrix} e^{\ii \phi/2} & 0 \\ 0 & e^{-\ii \phi/2} \end{pmatrix}.
\eeq
This is of the form $\rho =U^* D_2(\lambda_1,\lambda_2)U$ where the diagonal matrix of
the eigenvalues $D_2(\lambda_1,\lambda_2)$ is sandwiched between a unitary matrix $U$
and its adjoint $U^*$ when we set $x_i= \lambda_i$. This agrees with the parametrization
(\ref{eq:dens-matrix-param}) for $n=2$ with $$U=A_{2,2}= \begin{pmatrix}
 1-(1-c_2)|z_2\ran\lan z_2| \; & s_2 |z_2\ran \\ -s_2 \lan z_2| & c_2
\end{pmatrix}$$
when we set $z_2 \in S(\C)$ as $z_2 = e^{-\ii \phi}$ and $\theta_2 =\vartheta \in [0,\pi/2]$.

Knowing the explicit form of a density matrix $\rho_2$ in two dimension we can easily
calculate the explicit form of a density matrix on $\C^3$ by using the recursion
formula (\ref{rec-rel-dens}): Given any point
$(\lambda_1,\lambda_2,\lambda_3; \theta_2, z_2; \theta_3, z_3) \in Q_3$ we find
\begin{align} \label{3D-dens-mat}
\rho_3 &=A_{3,3}(\theta_3,z_3)^*A_{3,2}(\theta_2,z_2)^* D_3(\lambda_1,\lambda_2,\lambda_3) A_{3,2}(\theta_2,z_2)A_{3,3}(\theta_3,z_3)\\
&=A_{3,3}^* \begin{pmatrix}
\rho_2 \pr \; & 0\\ 0 & \lambda_3 \end{pmatrix} A_{3,3},
\end{align}
where $\rho_2 \pr$ is a positive matrix with trace $\Tr{\rho_2 \pr} = \lambda_1 + \lambda_2$.

Clearly, this recursive construction can be continued to $n=4,5, \ldots$.

\subsection{Jarlskog parametrization of density matrices for composite systems}
Consider now a density operator for the composite system living in $\C^n \otimes \C^m$.
It is clear that we may parametrize it as a density operator living in $\C^{nm}$.
However, by doing this we lose information about the particular tensor product
structure of the total Hilbert space $\C^{nm}$. To control the division into subsystems
of $\C^{nm}=\C^{n}\otimes \C^m$ let us consider $\rho$ as an $n \times n$ matrix with $m \times m$
blocks, i.e.,
\beq \label{comp-dens-m}
\rho_{n,m} =\sum_{i,j=1}^n |i\ran \lan j| \otimes \rho_{ij},  \eeq
with $\rho_{ij}$ being $m \times m$ complex matrices. Our aim is to provide a suitable
parametrization for positive block matrices. We proceed in analogy to the previous section.

For $\und{\lambda} = (\lambda_1, \lambda_2,\ldots,\lambda_{nm}) \in \Lambda_{nm} $ we denote
by $D_{nm}(\und{\lambda})$ the diagonal matrix of eigenvalues. Then for any $U_{n,m} \in SU(nm)$
\beq \label{d-mat}
\rho_{n,m}= U_{n,m}^* D_{nm}(\und{\lambda}) U_{n,m} \eeq
is a density matrix in $\C^{nm}$. Again we parametrize elements $U_{n,m} \in SU(nm)$ by
\beq \label{Unm-comp}
su(nm) \ni X_1 + X_2 + \cdots + X_n \longrightarrow e^{X_1}e^{X_2}\cdots e^{X_n} \in SU(nm) \eeq
where now $X_1$ is an $n \times$ anti-hermitian block-diagonal matrix with $m \times m$ blocks,
and for $j=2,\ldots, n$ $X_j$ is an $n \times$ anti-hermitian  matrix with $m \times m$ blocks defined as follows:
\beq \label{Xj-form}
X_j = \begin{pmatrix}
I_{j-1} \otimes \0_m \; & |Z_j \ran & \0 \\[2mm]
- \lan Z_j | & \0_m & \0 \\[2mm]
\0 & \0 & \; I_{n-j} \otimes \0_m
\end{pmatrix} ,\eeq
where instead of the $n-1$ column vectors $z_j$ used in the previous section we choose now $n-1$ column
block vectors
\beq \label{Zj-blocks}
|Z_j \ran = \begin{pmatrix}
Z_{1,j} \\ Z_{2,j} \\ \vdots\\ Z_{j-1,j} \end{pmatrix}, \eeq
with $Z_{i,j}$ being $m \times m$ matrices; and similarly
$$\lan Z_j| =( Z_{1,j}^*,Z_{2,j}^*,\ldots,Z_{j-1,j}^*) .$$
The Jarlskog matrices $A^j_{n,m} = e^{X_j}$ are now of the following form:
\beq \label{A1-Jarlskog}
A^1_{n,m} = \begin{pmatrix}
U_1 & \0_m & \cdots &\0_m \\[2mm]
\0_m & U_2 & \cdots  & \0_m \\[2mm]
\vdots & \vdots & \ddots & \vdots \\[2mm]
\0_m & \0_m & \cdots & U_n               \end{pmatrix},
\eeq
where the $U_k$ are unitary $m \times m$ matrices. For $j=2,\ldots,n$ one finds

\beq \label{Aj-Jarlskog}
A^j_{n,m} = \begin{pmatrix}
V^j_{n,m} & \0_m & \cdots &\0_m \\[2mm]
\0_m & I_m & \cdots  & \0_m \\[2mm]
\vdots & \vdots & \ddots & \vdots \\[2mm]
\0_m & \0_m & \cdots &  I_m               \end{pmatrix},
\eeq
where the $m$ dimensional unit matrix $I_m$ appears $n-j$ times and where $V^j_{n,m}$ is a unitary $ j \times j$ block matrix with $m \times m$ blocks, defined as follows:
\beq \label{Vjnm}
V^j_{n,m} = \begin{pmatrix}
 I_{j-1} \otimes I_m - |\tilde{Z}_j \ran [I_{j-1}\otimes (I_m -C_j)]\lan \tilde{Z}_j| \quad &
 \; |\tilde{Z}_j \ran S_j \\[2mm]
 -S_j \lan \tilde{Z}_j| \; & \; C_j
\end{pmatrix}.
\eeq
Here we use the following notation: $|\tilde{Z}_j \ran$ denotes the normalized block vector
\beq \label{b-vec}
\tilde{Z}_j = \frac{1}{\norm{Z_j}} Z_j,\quad \norm{Z_j}^2 = Z_{1,j}^* Z_{1,j}+ \cdots + Z_{j-1,j}^* Z_{j-1,j},
\eeq
and $ I_{j-1} \otimes I_m - |\tilde{Z}_j \ran [I_{j-1}\otimes (I_m -C_j)]\lan \tilde{Z}_j|$ stands for the
$(j-1)\times (j-1)$ block matrix
\beq \label{b-mat}
\sum_{k,l=1}^{j-1} |k\ran \lan l| \otimes \tilde{Z}_{k,l}^* C_j \tilde{Z}_{l,j},
\eeq
with
\beq \label{cs}
 C_j = \cos{\Xi_j}, \quad S_j = \sin{\Xi_j}, \quad \Xi_j =\norm{Z_j}.
\eeq
Our parametrization (\ref{d-mat}) of density matrices $\rho_{n,m}$ now reads
\begin{eqnarray}     \label{d-mat-par}
\rho_{n,m} &=&A_{n,m}^{n*}A_{n,m}^{n-1*} \cdots A_{n,m}^{1*}D_{nm}(\und{\lambda}) A_{n,m}^{1}\cdots A_{n,m}^{n-1}A_{n,m}^{n} \nonumber\\
&= &A_{n,m}^{n*} \cdots A_{n,m}^{2*}D_{n}(\Lambda_1|\cdots |\Lambda_n) A_{n,m}^{2}\cdots A_{n,m}^{n},
\end{eqnarray}
where
\beq \label{diag-block}
D_{n}(\Lambda_1|\cdots |\Lambda_n) =A_{n,m}^{1*}D_{nm}(\und{\lambda}) A_{n,m}^{1}
\eeq
is a positive block diagonal matrix with diagonal blocks
\beq \label{d-block}
\Lambda_k =U_k^* D(\lambda_{km}, \ldots, \lambda_{km +m -1})U_k
\eeq
which are positive $m \times m$ matrices and which satisfy the normalization condition
\beq \label{trace-normal}
\Tr{(\Lambda_1 + \cdots \Lambda_n)} =1 .
\eeq

\subsubsection{$2 \times 2$ systems}
Let us consider a $2 \otimes 2$ system to illustrate our parametrization for
the well-known 2-qubit states. Taking
\beq \label{choice1}
\Lambda_1 = \0_2,\quad \Lambda_2= \frac{1}{2}(I_2 -\sigma_z),\quad S= \sin{\alpha}\;I_2,\quad C=\cos{\alpha}\;I_2, \quad U=\sigma_z
\eeq
one obtains a family of rank-1 projectors
\beq \label{proj1}
P(\alpha) = \begin{pmatrix}
\sin^2{\alpha} & 0 \quad & 0 \;& \sin{\alpha} \cos{\alpha} \\
\sin{\alpha} \cos{\alpha}\; & 0 \quad & 0\; & 0\\[2mm]
0& 0 \quad & 0\; & 0\\
\sin{\alpha} \cos{\alpha}\; & 0\quad & 0\;& \cos^2{\alpha}
\end{pmatrix},
\eeq
which corresponds to a pure state
$$\psi_{\alpha} = \sin{\alpha}\; |00\ran + \cos{\alpha}\;|11\ran.$$
Note that this state is separable if and only if $S = 0$ or $C = 0$. For
$S = C = I_2/\sqrt{2}$ , one obtains a maximally entangled state. It shows that a nontrivial
rotation by $\alpha$ does produce quantum entanglement.

As a second example in
this class let us take $S = C = I_2/\sqrt{2}, \;U = \sigma_x$ and
\beq \label{Lambda12}
\Lambda_1 = \frac{1}{4} \begin{pmatrix} 1-p & 0 \\ 0 & 1-p \end{pmatrix}, \qquad
\Lambda_2 = \frac{1}{4} \begin{pmatrix} 1-p & 0 \\ 0 & 1+ 3p \end{pmatrix},
\eeq
with $-1/3 \leq p \leq 1$ to guarantee positivity of the matrices $\Lambda_i$. One obtains
the partially transposed Werner state (\ref{Wpt})
\beq \label{2-qubit1}
W_{\textrm{pt}}(-p) = \frac{1}{4} \begin{pmatrix}
1+p & 0 \quad & 0 & 2p \\ 0 & 1-p \quad & 0 & 0 \\[2mm] 0 & 0 \quad & 1-p & 0\\ 2p & 0 \quad & 0 & 1+p
\end{pmatrix}. \eeq
As has been shown earlier, this state is separable if and only if $p \leq 1/3$.
The point $p = 1/3$ is not distinguished by our parametrization.

Next we consider $S=\sin{\alpha}\;I_2$ and $C= \cos{\alpha}\; I_2$ and obtain a more general two-parameter
family
\beq \label{2param}
I(p,\alpha)=\frac{1}{4} I_2 \otimes I_2 + p P(\alpha)
\eeq
of $2 \otimes 2$ states. This family is separable if and only if
$$p \leq \frac{1}{1 + 2 \sin{(2 \alpha)}}.$$

This family of states can be generalized further as follows: Instead of (\ref{Lambda12}) we take
\beq \label{Lambda1234}
\Lambda_1 = \begin{pmatrix} p_2 & 0 \\0& p_4
\end{pmatrix}, \qquad \Lambda_2 = \begin{pmatrix} p_3 & 0 \\0& p_1
\end{pmatrix}, \eeq
where $p_i \geq 0$ and $p_1 + p_2 +p_3 +p_4 =1$. Furthermore, the $S$ and the $C$ matrices are chosen as
$$S= \begin{pmatrix} \sin{\alpha}&0\\0&\sin{\beta} \end{pmatrix},\quad C= \begin{pmatrix}
\cos{\alpha} & 0 \\ 0& \cos{\beta} \end{pmatrix},\qquad \alpha,\beta \in [0,\pi/2].$$ With $U=\sigma_x$
the following family of states results ($s_{\alpha}=\sin{\alpha},\;c_{\alpha}= \cos{\alpha}$ and similarly for $s_{\beta}$ and $c_{\beta}$, $\und{p} = (p_1,p_2,p_3,p_4)$ as above):
\begin{equation} \label{5parameter}
\rho(\und{p};\alpha,\beta) =
\begin{pmatrix}
p_1 c_{\alpha}^2 + p_2 s_{\alpha}^2\; & 0 \quad & 0\; & (p_1 -p_2) s_{\beta} c_{\beta} \\
0 \;&p_3 c_{\beta}^2 + p_4 s_{\beta}^2 \quad & (p_3 -p_4) s_{\alpha} c_{\alpha}\; &0\\[2mm]
0 \; & (p_3 -p_4) s_{\alpha} c_{\alpha} \quad & p_3 s_{\beta}^2 + p_4 c_{\beta}^2 & 0 \\
(p_1 -p_2) s_{\beta} c_{\beta} \; & 0 \quad & 0 \; &p_1 s_{\alpha}^2 + p_2 c_{\alpha}^2\
\end{pmatrix}\; .
\end{equation}
By construction or by a direct calculation we see that $\rho(\und{p};\alpha,\beta) \geq 0$ and $\Tr{\rho(\und{p};\alpha,\beta)}=1$ for any choice of the parameters with the restrictions given above.

Note that the above family belongs to the class of $2 \otimes 2$ circulant states considered in \cite{CK07}.
Note also that for $\alpha=\beta= \pi/4$ the family of states (\ref{5parameter}) reduces to the family of Bell diagonal states
\beq \label{Bell-diag-states}
\rho(\und{p}) = \frac{1}{2} \begin{pmatrix}
p_1 + p_2 \;& 0 \quad & 0\; & p_1 - p_2\\
0\; & p_3 + p_4 \quad &p_3 -p_4 \; & 0 \\[2mm]
0\;& p_3 -p_4 \quad & p_3 + p_4 \; & 0 \\
p_1 -p_2 \;& 0 \quad & 0 \;& p_1 +p_2
\end{pmatrix}\; .
\eeq
In \cite{CK07} the separability of these states has been investigated  for various values of the parameters.

\subsubsection{$2 \otimes m$ systems}
For $n=2$ and $m \geq 3$ our formula (\ref{d-mat-par}) reads
\beq  \label{2-m}
\rho_{2,m}={A^{2*}_{2,m}} D_2(\Lambda_1|\Lambda_2)A^2_{2,m},\quad A^2_{2,m}=V^2_{2,m}=
\begin{pmatrix} \tilde{Z}C \tilde{Z}^* \quad & \tilde{Z}S\\[1mm] -S\tilde{Z}^* \quad & C
\end{pmatrix}
\eeq
with $\tilde{Z}=U \in U(m)$ and $C = \cos{\Xi_2}$ and $S= \sin{\Xi_2}$.
$ D_2(\Lambda_1|\Lambda_2)$ is given by (\ref{diag-block}), (\ref{d-block}), and (\ref{trace-normal}).
Matrix multiplication gives
\beq \label{2-m-d-matrix}
\rho_{2,m} =\begin{pmatrix} U^* \;& \0_m \\ \0_m\;& I_m\end{pmatrix}\;
\begin{pmatrix} C U^* \Lambda_1 U C + S \Lambda_2 S \quad & S\Lambda_2 C -CU^* \Lambda_1 U S \\[2mm]
C\Lambda_2 S -SU^* \Lambda_1 U C \quad & C  \Lambda_2  C + Su^* \Lambda_1 u S\end{pmatrix}\;
\begin{pmatrix} U\;& \0_m \\ \0_m\;& I_m\end{pmatrix}
\eeq
By choosing particular values for the parameters appearing in (\ref{2-m-d-matrix}) we find some examples which have been considered in the literature.

For $S=\0$ or $C=\0$ one obtains a class of block diagonal matrices
\beq \label{block-diag}
\begin{pmatrix} \Lambda_1 \; & \0_m \\ \0_m \;& \Lambda_2 \end{pmatrix} \quad \rm{or} \quad
\begin{pmatrix} \Lambda_2 \; & \0_m \\ \0_m \;& \Lambda_1 \end{pmatrix}.
\eeq
These matrices represent separable $2 \otimes m$ states and thus show that quantum entanglement arises only
for nontrivial $\Xi_2$ such that $C\neq \0$ and $S \neq \0$.

Next consider the case $\Lambda_1 =\Lambda_2 = \Lambda$ and $[\Lambda,U]=0$. We get the following class of
$2 \otimes m$ states:
\beq \label{spec-block}
\begin{pmatrix} U^* \;& \0_m \\ \0_m\;& I_m\end{pmatrix}\;
\begin{pmatrix} A \;& B \\ B^* \;& A
\end{pmatrix}
\begin{pmatrix} U \;& \0_m \\ \0_m\;& I_m\end{pmatrix}\;
\eeq
with
$$A = C\Lambda C + S\Lambda S, \qquad B = S\Lambda C = C\Lambda S .$$
If in addition $U^* A U = A$ is assumed one gets the matrices
\beq \label{Toep-block}
 \begin{pmatrix} A \quad & UB \\ (UB)^* \; & A \end{pmatrix}.
\eeq
These are block Toeplitz positive matrices and it is well-known that they are separable \cite{GB03}.
Thus (\ref{Toep-block}) defines a huge family of bipartite separable states.

Similarly, the block Hankel positive matrices of \cite{GB03} can be reconstructed. To this end we assume that the matrices $U,\Lambda_1,\Lambda_2$ and $\Xi_2$ satisfy
$$[U^* \Lambda_1 U, \Xi_2]=0,\qquad [\Lambda_2,\Xi_2]=0.$$
This produces the following class of $2 \otimes m $ states:
\beq \label{spec-block2}
\begin{pmatrix} U^* \;& \0_m \\ \0_m\;& I_m\end{pmatrix}\;
\begin{pmatrix} A_1 \;& B\pr \\ B\pr \;& A_2
\end{pmatrix}
\begin{pmatrix} U \;& \0_m \\ \0_m\;& I_m\end{pmatrix}\;
\eeq
with $A_i = C\Lambda_i C + S\Lambda_i S$ and $B\pr = SC(\Lambda_2 - U^* \Lambda_1 U)$. Under the additional assumption $UB\pr= B\pr U$ we arrive at matrices of the form
\beq \label{Hank-block}
\begin{pmatrix} U^* A_1 U\; & X  \\ X \; & A_2 \end{pmatrix}
\eeq
with $X=UB\pr$. These are block Hankel positive matrices and hence separable \cite{GB03}.

\section{Conclusion}\label{sec:Conclusion}
In this review we have discussed different parametrizations of $n\times n$ density matrices.
We have compared three different parametrizations, namely the Bloch vector, coset, and Jarlskog parametrizations. Of these the Bloch vector parametrization is the oldest and most widely used,
while the two others are relatively recent discoveries and consequently not so well known.

The Bloch vector parametrization has found more applications in physical problems than any other
density matrix parametrization. This is due to the inherent simplicity of the parametrization. The basis matrices used in this representation are hermitian, which guarantees that the components of the
Bloch vector are real. Furthermore, these components can be obtained straightforwardly  as expectation values of hermitian operators.
Therefore, given a density operator, the calculation of the components
of the corresponding Bloch vector can be done straightforwardly.
On the other hand, there is a serious disadvantage in the Bloch vector parametrization:
In  all the cases of $n\geq 3$ it is practically impossible to determine the parameter set corresponding to physical states  explicitly (see Section 2.2). Only for a two-level system the parameter set is easily determined
and is given by unit ball in $\mathbb{R}^3$, the so called Bloch ball.
The difficulty derives from the requirement that a density matrix has to be a positive operator. As yet, there is no simple way to determine which Bloch vectors lead to positive matrices. 
Despite the complexity of the parameter set, the Bloch vector parametrization has many applications. It has been known for a long time that it gives a bijective mapping between the states of a two-level system and the points of the Bloch ball, providing an elegant
way to visualize the states and dynamics of two-level systems.
Another advantage is that the hermiticity of the basis enables to write the density matrices and Hamiltonian in the same basis. This makes possible to derive easily the differential equation giving the time-evolution of the Bloch vector. Under some conditions for the Hamiltonian of the system, this approach allows to identify various constants of motion. The Bloch vector has turned out to be useful also when the entanglement and separability of two-qubit states is examined.

The second parametrization discussed in detail in this article is the Jarlskog parametrization which is based on a suitable parametrization of (special) unitary matrices. Its main advantages are (see sections 4.2 and 5.1):
\begin{itemize}
\item Its parameter set is given explicitly;
\item it is recursive;
\item it extends naturally to composite systems.
\end{itemize}
However,  in the form presented here,  it contains redundancy
in the case of a degenerate spectrum. We have indicated how to eliminate this and get injectivity for this parametrization, too.

\bibliographystyle{tMOP}
\bibliography{d-matrix}
\end{document}